\documentclass[a4paper,ngerman,magyar,english]{article}
\usepackage{mathpazo}
\usepackage[T1]{fontenc}
\usepackage[latin2,latin9]{inputenc}
\usepackage{units}
\usepackage{amsthm}
\usepackage{amsmath}
\usepackage{amssymb}
\usepackage{graphicx}

\makeatletter

\pdfpageheight\paperheight
\pdfpagewidth\paperwidth

\newenvironment{lyxlist}[1]
{\begin{list}{}
{\settowidth{\labelwidth}{#1}
 \setlength{\leftmargin}{\labelwidth}
 \addtolength{\leftmargin}{\labelsep}
 }}
{\end{list}}
\theoremstyle{plain}
\newtheorem{thm}{\protect\theoremname}

\allowdisplaybreaks[0]

\newcounter{meg}

\newcommand{\R}{%
\refstepcounter{meg}%
{\vskip6pt \noindent \bf Remark \themeg.\ \ \ }%
}

\makeatother

\usepackage{babel}
  \addto\captionsenglish{\renewcommand{\theoremname}{Theorem}}
  \addto\captionsmagyar{\renewcommand{\theoremname}{Tétel}}
  \addto\captionsngerman{\renewcommand{\theoremname}{Theorem}}
\addto\captionsenglish{}
\addto\captionsmagyar{}
\addto\captionsngerman{}
\providecommand{\theoremname}{Theorem}

\begin{document}

\title{Is the relativity principle consistent with classical electrodynamics?\textbf{}\\
\textbf{ }{\normalsize Towards a logico-empiricist reconstruction
of a physical theory}}

\author{Márton Gömöri and László E. Szabó\emph{\small }\\
\emph{\small Department of Logic, Institute of Philosophy}\\
\emph{\small Eötvös University, Budapest}\\
\emph{\small http://phil.elte.hu/logic}}

\date{{\normalsize ~}}
\maketitle
\begin{abstract}
It is common in the literature on classical electrodynamics (ED) and
relativity theory that the transformation rules for the basic electrodynamical
quantities are derived from the hypothesis that the relativity principle
(RP) applies to Maxwell's electrodynamics. As it will turn out from
our analysis, these derivations raise several problems, and certain
steps are logically questionable. This is, however, not our main concern
in this paper. Even if these derivations were completely correct,
they leave open the following questions: (1)~Is the RP a true law
of nature for electrodynamical phenomena? (2)~Are, at least, the
transformation rules of the fundamental electrodynamical quantities,
derived from the RP, true? (3)~Is the RP consistent with the laws
of ED in a single inertial frame of reference? (4)~Are, at least,
the derived transformation rules consistent with the laws of ED in
a single frame of reference? Obviously, (1) and (2) are empirical
questions. In this paper, we will investigate problems (3) and (4). 

First we will give a general mathematical formulation of the RP. In
the second part, we will deal with the operational definitions of
the fundamental electrodynamical quantities. As we will see, these
semantic issues are not as trivial as one might think. In the third
part of the paper, applying what J. S. Bell calls \textquotedblleft{}Lorentzian
pedagogy\textquotedblright{}---according to which the laws of physics
in any one reference frame account for all physical phenomena---we
will show that the transformation rules of the electrodynamical quantities
are identical with the ones obtained by presuming the covariance of
the equations of ED, and that the covariance is indeed satisfied.

As to problem (3), the situation is more complex. The covariance of
the physical equations is actually not enough for the RP; whether
the RP holds depends on the details of the solutions describing moving
objects. As we will see, in case of ED, the very concept of a moving
system raises serious conceptual problems. Thus, contrary to the widespread
views, we will conclude that the question asked in the title has no
obvious answer.

\tableofcontents{}
\end{abstract}

\section{Introduction\label{sec:Introduction}}

It is common in the literature on classical electrodynamics (ED) and
relativity theory that the transformation rules for the basic electrodynamical
quantities are \emph{derived} from the assumption that the relativity
principle (RP) applies to Maxwell's electrodynamics. As it will turn
out from our analysis (the details are given in Remark~\ref{meg:7}
and \ref{meg:8}), these derivations raise several problems, and certain
steps are logically questionable. This is, however, not our main concern
in this paper. Even if these derivations were completely correct,
they leave open the following questions: 
\begin{lyxlist}{00.00.0000}
\item [{(Q1)}] Is RP a true law of nature for electrodynamical phenomena? 
\item [{(Q2)}] Are, at least, the transformation rules of the fundamental
electrodynamical quantities, derived from RP, true? 
\end{lyxlist}
First of all, one has to clarify what the principle says. The RP is
usually formulated as follows: {}``All the laws of physics take the
same form in any inertial frame of reference.'' This short sentence,
however, does not express exactly what the principle actually asserts.
For example, consider how Einstein applies the principle in his 1905
paper:
\begin{quotation}
Let there be given a\emph{ stationary} rigid rod; and let its length
be $l$ as \emph{measured by a measuring-rod which is also stationary}.
We now imagine the axis of the rod lying along the axis of $x$ of
the stationary system of co-ordinates, and that a\emph{ }uniform\emph{
}motion of parallel translation with velocity $v$ along the axis
of $x$ in the direction of increasing $x$ is then imparted to the
rod. We now inquire as to the length of the \emph{moving} rod, and
imagine its length to be ascertained by the following \emph{two} operations:
\begin{lyxlist}{0.0.0}
\item [{(a)}] The observer \emph{moves together with the given measuring-rod
and the rod to be measured}, and \emph{measures the length} of the
rod directly by superposing the measuring-rod, \emph{in just the same
way} \emph{as if all three were at rest}.
\item [{(b)}] By means of \emph{stationary} clocks set up in the \emph{stationary}
system and synchronizing in accordance with {[}the light-signal synchronization{]},
the observer ascertains at what points of the \emph{stationary} system
the two ends of the rod to be measured are located at a definite time.
The distance between these two points, measured by the measuring-rod
already employed, which in this case is \emph{at rest}, is also a
length which may be designated {}``the length of the rod.'' 
\end{lyxlist}
\noindent \emph{In accordance with the principle of relativity} the
length \emph{to be discovered by the operation (a)}---we will call
it {}``the length of the rod in the moving system''---must be equal
to the length $l$ of \emph{the stationary} rod. 

The length to be discovered by the operation (b) we will call {}``the
length of the (\emph{moving}) rod in the \emph{stationary} system.''
This we shall determine on the basis of our two \emph{principles},
and we shall find that it \emph{differs} from $l$. {[}all italics
added{]}
\end{quotation}
From a careful reading of this simple example of Einstein, and also
from other usual applications of the RP, for example from the usual
derivation of the electromagnetic field of a uniformly moving point
charge (Remark~\ref{meg:3a}), one concludes with the following more
detailed formulation (Szabó~2004):
\begin{lyxlist}{00.00.0000}
\item [{(RP)}] The physical description of the behavior of a system co-moving
as a whole with an inertial frame\emph{ $K$}, expressed in terms
of the results of measurements obtainable by means of measuring equipments
co-moving with\emph{ $K$, }takes the same form as the description
of the similar behavior of the same system when it is co-moving with
another inertial frame\emph{ $K'$}, expressed in terms of the measurements
with the same equipments when they are co-moving with\emph{ $K'$}.\emph{ }
\end{lyxlist}
(Q1) is a legitime question, in spite of the obvious fact that the
RP is a meta-law, that is a law about the laws of nature. For, whether
it is true or not is determined by the laws of nature; whether the
laws of nature are true or not depends on how the things are in the
physical world. So, in spite of the formal differences, the epistemological
status of the RP is ultimately the same as that of the ordinary physical
laws.

Apparently, to answer question (Q1), that is to verify whether the
principle holds for the laws describing electromagnetic phenomena,
the following will be needed:
\begin{lyxlist}{00.00.0000}
\item [{(a)}] We must be able to tell when two electrodynamical systems
are the same except that they are moving, as a whole, relative to
each other---one system is co-moving with $K$, the other is co-moving
with $K'$. 
\item [{(b)}] We must have proper descriptions of the behavior of both
systems, expressed in terms of two different sets of corresponding
variables---one belonging to $K$ the other to $K'$.
\item [{(c)}] The RP would be completely meaningless if we mix up different
physical quantities, because, in terms of different variables, one
and the same physical law in one and the same inertial frame of reference
can be expressed in different forms. Consequently, we must be able
to tell which variable in $K$ corresponds to which variable in $K'$;
that is, how the physical quantities defined in the two different
inertial frames are identified. Also, question (Q2) by itself would
be meaningless without such an identification. The most obvious idea
is that we identify those physical quantities that have identical
empirical definitions. 
\item [{(d)}] The empirical definition of a physical quantity is based
on standard measuring equipments and standard operational procedures.
How do the observers in different reference frames share these standard
measuring equipments and operational procedures? Do they all base
their definitions on the same standard measuring equipments? On the
one hand, they must do something like that, otherwise any comparison
between their observations would be meaningless. On the other hand,
however, it is quite obvious that the principle is understood in a
different way---see the above quoted passage of Einstein. That is
to say, if the standard measuring equipment defining a physical quantity
$\xi$ is, for example, at rest in $K$ and, therefore, moving in
$K'$, then the observer in $K'$ does not define the corresponding
$\xi'$ as the physical quantity obtainable by means of the original
standard equipment---being at rest in $K$ and moving in $K'$---but
rather as the one obtainable by means of the same standard equipment
\emph{in another state of motion}, namely when it is at rest in $K'$
and moving in $K$. Thus, we must be able to tell when two measuring
equipments are the same, except that they are moving, as a whole,
relative to each other---one is at rest relative to $K$, the other
is at rest relative to $K'$. Similarly, we must be able to tell when
two operational procedures performed by the two observers are the
{}``same''; in spite of the fact that the procedure performed in
$K'$ obviously differs from the one performed in $K$.
\item [{(e)}] Obviously, in order to compare these procedures we must know
what the procedures exactly are; that is, we must have precise operational
definitions of the quantities in question. 
\end{lyxlist}
All these issues naturally arise if we want to verify \emph{empirically}
whether the RP is a true law of nature for electrodynamical phenomena.
For, empirical verification, no doubt, requires the physicist to know
which body of observational data indicates that statement (RP) is
true or false. Without entering here into the discussion of verificationism
in general, we have only two remarks to make. 

First, our approach is entirely compatible with confirmation/semantic
holism. The position we are advocating here is essentially holistic.
We accept it as true that {}``our statements about the external world
face the tribunal of sense experience not individually but only as
a corporate body'' (Quine 1951). On the one hand this means that
a theory, together with its semantics, as a whole is falsified if
any single sentence of its deductive closure is empirically falsified;
any part of the theory can be reconsidered---the basic deductive system,
the applied mathematical tools, and the semantic rules of correspondence
included. On the other hand, contrary to what is often claimed, this
kind of holism does not imply that the sentences of a physical theory,
at least partly, cannot be provided with empirical meaning by reducing
them to a sense-datum language. In our view, on the contrary, what
semantic holism implies is that the empirical definition of a physical
term must not be regarded in isolation from the empirical definitions
of the other terms involved in the definition. For example, as we
will see, the empirical definitions of electrodynamical quantities
cannot be separated from the notion of mass; in fact, the definitions
in the usual ED and mechanics textbooks, together, constitute an incoherent
body of definitions with circularities. This is perhaps a forgivable
sin in the textbook literature. But, in philosophy of physics, the
recognition of these incoherencies should not lead us to jettison
the empirical content of an individual statement; on the contrary,
we have to reconstruct our theories on the basis of a sufficiently
large coherent body of empirical/operational definitions. In our understanding,
this is the real holistic approach---a super-holistic, if you like.

Second, in fact, our arguments in this paper will rely on the verificationist
theory of meaning in the following very weak sense: In physics, the
meaning of a term standing for a \emph{measurable quantity} which
is supposed to characterize an objective feature of physical reality
is determined by the empirical operations with which the value of
the quantity in question can be ascertained. Such a limited verificationism
is widely accepted among physicists; almost all ED textbooks start
with some descriptions of how the basic quantities like electric charge,
electric and magnetic field strengths, etc. are empirically interpreted.
Our concern is that these empirical definitions do not satisfy the
standard of the above mentioned super-holistic coherence, and the
solution of the problem is not entirely trivial.

In any event, in this paper, the demand for precise operational definitions
of electrodynamical quantities emerges not from this epistemological
context; not from philosophical ideas about the relationship between
physical theories, sense-data, and the external reality; not from
the context of questions (Q1) and (Q2). The problem of operational
definitions is raised as a problem of pure theoretical physics, in
the context of the \emph{inner consistency} of our theories. The reason
is that instead of the empirical questions (Q1) and (Q2) we will in
fact investigate the following two \emph{theoretical} questions:
\begin{lyxlist}{00.00.0000}
\item [{(Q3)}] Is the RP consistent with the laws of ED in a single inertial
frame of reference? 
\item [{(Q4)}] Are, at least, the derived transformation rules consistent
with the laws of ED in a single frame of reference? 
\end{lyxlist}
The basic idea is what J. S. Bell (1987, p. 77) calls {}``Lorentzian
pedagogy'', according to which {}``the laws of physics in any one
reference frame account for all physical phenomena, including the
observations of moving observers''. That is to say, if our physical
theories in any one reference frame provide a complete enough account
for our world, then all we will say about {}``operational'' definitions
and about {}``empirical'' facts---issues (a)--(e) included---must
be represented and accounted \emph{within the theory itself}; and
the laws of physics---again, in any one reference frame---must determine
whether the RP is true or not.

Thus, accordingly, the paper will consists of the following major
parts. First of all we will give a general mathematical formulation
of the RP and covariance. In the second part, we will clarify the
semantic issues addressed in point (e). In the third part, we will
derive the transformation rules of the electrodynamical quantities,
from the operational definitions and from the laws of ED in one inertial
frame of reference---independently of the RP; by which we will answer
our question (Q4). In this way---again, independently of the RP---we
will show the covariance of the Maxwell--Lorentz equations.

Whether the RP holds hinges on the details of the solutions describing
the behavior of moving objects. As we will see in the last section,
in case of ED, the very concept of a moving system raises serious
conceptual problems. Thus, contrary to the widespread views, we will
conclude that the question asked in the title has no obvious answer.

\medskip{}

\begin{center}
{*}{*}{*}\medskip{}

\par\end{center}

Throughout it will be assumed that space and time coordinates are
already defined in all inertial frames of reference; that is, in an
arbitrary inertial frame $K$, space tags $\mathbf{r}\left(A\right)=\left(x\left(A\right),y\left(A\right),z\left(A\right)\right)\in\mathbb{R}^{3}$
and a time tag $t\left(A\right)\in\mathbb{R}$ are assigned to every
event $A$---by means of some empirical operations.%
\footnote{In fact, to give precise empirical definitions of the basic spatio-temporal
quantities in physics is not a trivial problem (Szabó 2009).%
} We also assume that the assignment is mutually unambiguous, such
that there is a one to one correspondence between the space and time
tags in arbitrary two inertial frames of reference $K$ and $K'$;
that is, the tags $\left(x'\left(A\right),y'\left(A\right),z'\left(A\right),t'\left(A\right)\right)$
can be expressed by the tags $\left(x\left(A\right),y\left(A\right),z\left(A\right),t\left(A\right)\right)$,
and vice versa. The concrete form of this functional relation is an
empirical question. In this paper, we will take it for granted that
this functional relation is the well-known Lorentz transformation;
and the calculations, particularly in section~\ref{sec:Observations-of-moving},
will rest heavily on this assumption. It must be emphasized however
that we stipulate the Lorentz transformation of the kinematical quantities
as an \emph{empirical} fact, without suggesting that the usual \emph{derivations}
of these transformation rules from the RP/constancy of the speed of
light are unproblematic. In fact, these derivations raise questions
similar to (Q1)--(Q4), concerning the kinematical quantities. In this
paper, however, we focus our analysis only on the electrodynamical
quantities. 

It must be also noted that the transformation of the kinematical quantities,
alone, does not determine the transformation of the electrodynamical
quantities. As we will see, the latter is determined by the kinematical
Lorentz transformation in conjunction with the operational definitions
of the electrodynamical quantities and some empirical facts, first
of all the relativistic version of the Lorentz equation of motion.

Below we recall the most important formulas we will use. For the sake
of simplicity, we will assume the standard situation: the corresponding
axises are parallel and $K'$ is moving along the $x$-axis with velocity
$\mathbf{V}=\left(V,0,0\right)$ relative to $K$, and the two origins
coincide at time $0$.%
\footnote{All {}``vectors'' are meant to be in $\mathbb{R}^{3}$; boldface
letters $\mathbf{r},\mathbf{v},\mathbf{E}\ldots$ simply denote vector
matrices. %
} Throughout the paper we will use the following notations: $\gamma(\ldots)=\left(1-\frac{(\ldots)^{2}}{c^{2}}\right)^{-\frac{1}{2}}$
and $\gamma=\gamma(V)$.

The connection between the space and time tags of an event $A$ in
$K$ and $K'$ is the following:

\begin{eqnarray}
x'\left(A\right) & = & \gamma\left(x\left(A\right)-Vt\left(A\right)\right)\label{eq:LT1}\\
y'\left(A\right) & = & y\left(A\right)\\
z'\left(A\right) & = & z\left(A\right)\\
t'\left(A\right) & = & \gamma\left(t\left(A\right)-c^{-2}Vx\left(A\right)\right)\label{eq:LT4}
\end{eqnarray}
Let $A$ be an event on the worldline of a particle. For the velocity
of the particle at $A$ we have:

\noindent 
\begin{eqnarray}
v_{x}'\left(A\right) & = & \frac{v_{x}\left(A\right)-V}{1-c^{-2}v_{x}\left(A\right)V}\label{eq:LT5}\\
v_{y}'\left(A\right) & = & \frac{\gamma^{-1}v_{y}\left(A\right)}{1-c^{-2}v_{x}\left(A\right)V}\\
v_{z}'\left(A\right) & = & \frac{\gamma^{-1}v_{z}\left(A\right)}{1-c^{-2}v_{x}\left(A\right)V}\label{eq:LT8}
\end{eqnarray}
We shall use the inverse transformation in the following special case:
\begin{eqnarray}
\mathbf{v}'\left(A\right)=\left(v',0,0\right) & \mapsto & \mathbf{v}\left(A\right)=\left(\frac{v'+V}{1+c^{-2}v'V},0,0\right)\label{eq:sebesseg3}\\
\mathbf{v}'\left(A\right)=\left(0,0,v'\right) & \mapsto & \mathbf{v}\left(A\right)=\left(V,0,\gamma v'\right)\label{eq:sebesseg1}
\end{eqnarray}

\noindent The transformation rule of acceleration is much more complex,
but we need it only for $\mathbf{v}'\left(A\right)=\left(0,0,0\right)$:

\noindent 
\begin{eqnarray}
a'_{x}\left(A\right) & = & \gamma^{3}a_{x}\left(A\right)\label{eq:gyorsulas1}\\
a'_{y}\left(A\right) & = & \gamma^{2}a_{y}\left(A\right)\label{eq:gyorsulas2}\\
a'_{z}\left(A\right) & = & \gamma^{2}a_{z}\left(A\right)\label{eq:gyorsulas3}
\end{eqnarray}
We will also need the $y$-component of acceleration in case of $\mathbf{v}'\left(A\right)=\left(0,0,v'\right)$:
\begin{equation}
a'_{y}\left(A\right)=\gamma^{2}a_{y}\left(A\right)\label{eq:gyorsulas4}
\end{equation}

\section{The precise statement of the relativity principle\label{sec:Mathematics-of-relativity}}

In this section we will unpack the verbal statement (RP) and will
provide its precise mathematical formulation. Consider an arbitrary
collection of physical quantities $\xi_{1},\xi_{2},\ldots\xi_{n}$
in $K$, operationally defined by means of some\emph{ }operations
with some\emph{ }equipments being at rest in $K$. Let $\xi'_{1},\xi'_{2},\ldots\xi'_{n}$
denote another collection of physical quantities that are defined
by the \emph{same} \emph{operations} with the \emph{same} \emph{equipments},
but \emph{in different state of motion}, namely, in which they are
all moving with velocity $\mathbf{V}$ relative to $K$, that is,
co-moving with\emph{ $K'$}. Since, for all $i=1,2,\ldots n$, both
$\xi_{i}$ and $\xi'_{i}$ are measured by the same equipment---although
in different physical conditions---with the same pointer scale, it
is plausible to assume that the possible values of $\xi_{i}$ and
$\xi'_{i}$ range over the same $\sigma_{i}\subseteq\mathbb{R}$.
We introduce the following notation: $\Sigma={\displaystyle \times_{i=1}^{n}\sigma_{i}}\subseteq\mathbb{R}^{n}$.

It must be emphasized that quantities $\xi_{1},\xi_{2},\ldots\xi_{n}$
and $\xi'_{1},\xi'_{2},\ldots\xi'_{n}$ are, a priori, \emph{different}
physical quantities, due to the fact that the operations by which
the quantities are defined are performed under \emph{different} physical
conditions; with measuring equipments of \emph{different} states of
motion. Any objective (non-conventional) relationship between them
must be a contingent law of nature (also see Remark~\ref{meg:5}).
Thus, the same numeric values, say, $(5,12,\ldots61)\in\mathbb{R}^{n}$
correspond to different states of affairs when $\xi_{1}=5,\xi_{2}=12,\ldots\xi_{n}=61$
versus $\xi'_{1}=5,\xi'_{2}=12,\ldots\xi'_{n}=61$. Consequently,
$\left(\xi_{1},\xi_{2},\ldots\xi_{n}\right)$ and $\left(\xi'_{1},\xi'_{2},\ldots\xi'_{n}\right)$
are \emph{not} elements of the \emph{same }{}``space of physical
quantities''; although the numeric values of the physical quantities,
in both cases, can be represented in $\Sigma={\displaystyle \times_{i=1}^{n}\sigma_{i}}\subseteq\mathbb{R}^{n}$.

Mathematically, one can express this fact by means of two different
$n$-dimensional manifolds, $\Omega$ and $\Omega'$, each covered
by one global coordinate system, $\phi$ and $\phi'$ respectively,
such that $\phi:\Omega\rightarrow\Sigma$ assigns to every point of
$\Omega$ one of the possible $n$-tuples of numerical values of physical
quantities $\xi_{1},\xi_{2},\ldots\xi_{n}$ and $\phi':\Omega'\rightarrow\Sigma$
assigns to every point of $\Omega'$ one of the possible $n$-tuples
of numerical values of physical quantities $\xi'_{1},\xi'_{2},\ldots\xi'_{n}$
\begin{figure}
\begin{centering}
\includegraphics[width=0.7\columnwidth]{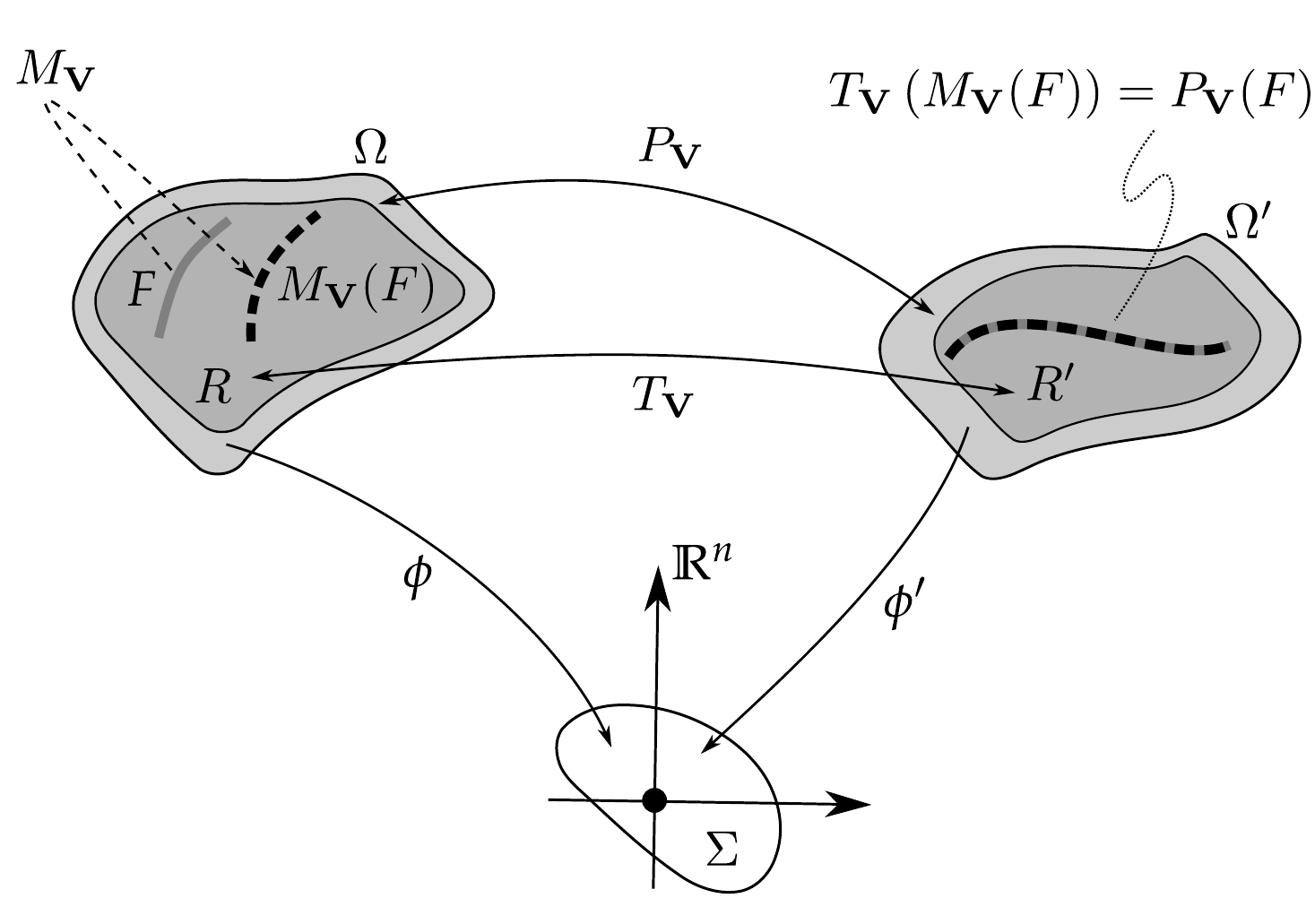}
\par\end{centering}

\caption{The relativity principle}
\label{Flo:relativity}
\end{figure}
 (Fig.~\ref{Flo:relativity}). In this way, a point $\omega\in\Omega$
represents the \emph{class} of physical constellations characterized
by $\xi_{1}=\phi_{1}(\omega),\xi_{2}=\phi_{2}(\omega),\ldots\xi_{n}=\phi_{n}(\omega)$;
similarly, a point $\omega'\in\Omega'$ represents the physical constellation
characterized by $\xi'_{1}=\phi'_{1}(\omega'),\xi'_{2}=\phi'_{2}(\omega'),\ldots\xi'_{n}=\phi'_{n}(\omega')$.%
\footnote{$\phi_{i}=\pi_{i}\circ\phi$, where $\pi_{i}$ is the $i$-th coordinate
projection in $\mathbb{R}^{n}$.%
} Again, these physical constellations are generally different, even
in case of $\phi(\omega)=\phi'(\omega')\in\mathbb{R}^{n}$.

In the above sense, the points of $\Omega$ and the points of $\Omega'$
range over all possible value combinations of physical quantities
$\xi_{1},\xi_{2},\ldots\xi_{n}$ and $\xi'_{1},\xi'_{2},\ldots\xi'_{n}$.
It might be the case however that some combinations are impossible,
in the sense that they never come to existence in the physical world.
Let us denote by $R\subseteq\Omega$ and $R'\subseteq\Omega'$ the
physically admissible parts of $\Omega$ and $\Omega'$. Note that
$\phi(R)$ is not necessarily identical with $\phi'(R')$.%
\footnote{One can show however that $\phi(R)=\phi'(R')$ if the RP, that is
(\ref{eq:RP-math1}), holds.%
}

We shall use a bijection $P_{\mathbf{V}}:\Omega\rightarrow\Omega'$
({}``putting primes''; Bell 1987, p.~73) defined by means of the
two coordinate maps $\phi$ and $\phi'$: 
\begin{equation}
P_{\mathbf{V}}\overset{^{def}}{=}\left(\phi'\right)^{-1}\circ\phi\label{eq:F'}
\end{equation}

In contrast with $P_{V}$, we now introduce the concept of what we
call the {}``transformation'' of physical quantities. It is conceived
as a bijection 
\begin{equation}
T_{\mathbf{V}}:\Omega\supseteq R\rightarrow R'\subseteq\Omega'
\end{equation}
determined by the contingent fact that whenever a physical constellation
belongs to the class represented by some $\omega\in R$ then it also
belongs to the class represented by $T_{\mathbf{V}}(\omega)\in R'$,
and vice versa. Since $\xi_{1},\xi_{2},\ldots\xi_{n}$ can be various
physical quantities in the various contexts, nothing guarantees that
such a bijection exists. In this section, however, we will assume
the existence of $T_{V}$.

\R \label{meg:peldak}It is worthwhile to consider several examples.
\begin{lyxlist}{00.00.0000}
\item [{(a)}] Let $\left(\xi_{1},\xi_{2}\right)$ be $\left(p,T\right)$,
the pressure and the temperature of a given (equilibrium) gas; and
let $\left(\xi'_{1},\xi'_{2}\right)$ be $\left(p',T'\right)$, the
pressure and the temperature of the same gas, measured by the moving
observer in $K'$. In this case, there exists a one-to-one $T_{V}$:
\begin{eqnarray}
p' & = & p\label{eq:nyomas}\\
T' & = & T\gamma^{-1}\label{eq:homerseklet}
\end{eqnarray}
(Tolman 1949, pp. 158--159).%
\footnote{There is a debate over the proper transformation rules (Georgieu 1969).%
}~A point $\omega\in\Omega$ of coordinates, say, $p=101325$ and
$T=300$ (in units $Pa$ and $^{\circ}K$) represents the class of
physical constellations---the class of possible worlds---in which
the gas in question has pressure of $101325\, Pa$ and temperature
of $300\,{}^{\circ}K$. Due to (\ref{eq:homerseklet}), this class
of physical constellations is different from the one represented by
$P_{V}\left(\omega\right)\in\Omega'$ of coordinates $p'=101325$
and $T'=300$; but it is identical to the class of constellations
represented by $T_{V}\left(\omega\right)\in\Omega'$ of coordinates
$p'=101325$ and $T'=300\gamma^{-1}$.
\item [{(b)}] Let $\left(\xi_{1},\xi_{2},\ldots\xi_{10}\right)$ be $\left(t,x,y,z,E_{x},E_{y},E_{z},r_{x},r_{y},r_{z}\right)$,
the time, the space coordinates where the electric field strength
is taken, the three components of the field strength, and the space
coordinates of a particle. And let $\left(\xi'_{1},\xi'_{2},\ldots\xi'_{10}\right)$
be $\left(t',x',y',z',E'_{x},E'_{y},E'_{z},r'_{x},r'_{y},r'_{z}\right)$,
the similar quantities obtainable by means of measuring equipments
co-moving with $K'$. In this case, there is no suitable one-to-one
$T_{V}$, as the electric field strength in $K$ does not determine
the electric field strength in $K'$, and vice versa.
\item [{(c)}] Let $\left(\xi_{1},\xi_{2},\ldots\xi_{13}\right)$ be $\left(t,x,y,z,E_{x},E_{y},E_{z},B_{x},B_{y},B_{z},r_{x},r_{y},r_{z}\right)$
and let $\left(\xi'_{1},\xi'_{2},\ldots\xi'_{13}\right)$ be $\left(t',x',y',z',E'_{x},E'_{y},E'_{z},B'_{x},B'_{y},B'_{z},r'_{x},r'_{y},r'_{z}\right)$,
where $B_{x},B_{y},B_{z}$ and $B'_{x},B'_{y},B'_{z}$ are the magnetic
field strengths in $K$ and $K'$. In this case, in contrast with
(b), the well known Lorentz transformations of the spatio-temporal
coordinates and the electric and magnetic field strengths constitute
a proper one-to-one $T_{V}$.\hfill{} $\lrcorner$\medskip{}

\end{lyxlist}
\noindent Next we turn to the general formulation of the concept of
the \emph{description of a particular behavior} of a physical system,
say, in $K$. We are probably not far from the truth if we assume
that such a description is, in its most abstract sense, a \emph{relation}
between physical quantities $\xi_{1},\xi_{2},\ldots\xi_{n}$; in other
words, it can be given as a subset $F\subset R$. 

\R Consider the above example~(a) in Remark~\ref{meg:peldak}.
An isochoric process of the gas can be described by the subset $F$
that is, in coordinates, determined by the following single equation:
\begin{equation}
F\,\,\,\left\{ p=\kappa T\right.\label{eq:kappa}
\end{equation}
with a certain constant $\kappa$. 

To give another example, consider the case (b). The relation $F$
given by equations 
\begin{equation}
F\,\,\,\left\{ \begin{alignedat}{1}E_{x} & =E_{0}\\
E_{y} & =0\\
E_{z} & =0\\
r_{x} & =x_{0}+v_{0}t\\
r_{y} & =0\\
r_{z} & =0
\end{alignedat}
\right.
\end{equation}
with some specific values of $E_{0},x_{0},v_{0}$ describes a neutral
particle moving with constant velocity in a static homogeneous electric
field. \hfill{} $\lrcorner$\medskip{}

\noindent Of course, one may not assume that an arbitrary relation
$F\subset R$ has physical meaning. Let $\mathcal{E}\subset2^{R}$
be the set of those $F\subset R$ which describe a particular behavior
of the system. We shall call $\mathcal{E}$ the\emph{ set of equations}
describing the physical system in question. The term is entirely justified.
In practical calculations, two systems of equations are regarded to
be equivalent if and only if they have the same solutions. Therefore,
a system of equations can be identified with the set of its solutions.
In general, the equations can be algebraic equations, ordinary and
partial integro-differential equations, linear and nonlinear, whatever.
So, in its most abstract sense, a system of equations is a set of
subsets of $R$. 

Now, consider the following subsets%
\footnote{We denote the map of type $\Omega\rightarrow\Omega'$ and its direct
image maps of type $2^{\Omega}\rightarrow2^{\Omega'}$ and $2^{2^{\Omega}}\rightarrow2^{2^{\Omega'}}$
or their restrictions by the same symbol. %
} of $\Omega'$, determined by an $F\in\mathcal{E}$:
\begin{lyxlist}{0000.0000.00}
\item [{$P_{\mathbf{V}}(F)\subseteq\Omega'$}] which formally is the {}``primed
$F$'', that is a relation of exactly the same {}``form'' as $F$,
but in the primed variables $\xi'_{1},\xi'_{2},\ldots\xi'_{n}$. Note
that relation $P_{\mathbf{V}}(F)$ does not necessarily describe a
true physical situation, as it can be not realized in nature.
\item [{$T_{\mathbf{V}}(F)\subseteq R'$}] which is the same description
of the same physical situation as $F$, but \emph{expressed} in the
primed variables.
\end{lyxlist}
We need one more concept. The RP is about the connection between two
situations: one is in which the system, as a whole, is at rest relative
to inertial frame $K$, the other is in which the system shows the
similar behavior, but being in a collective motion relative to $K$,
co-moving with $K'$. In other words, we assume the existence of a
map $M_{\mathbf{V}}:\,\mathcal{E}\rightarrow\mathcal{E}$, assigning
to each $F\in\mathcal{E}$, stipulated to describe the situation in
which the system is co-moving as a whole with inertial frame $K$,
another relation $M_{\mathbf{V}}(F)\in\mathcal{E}$, describing the
similar behavior of the same system when it is, as a whole, co-moving
with inertial frame\emph{ $K'$}, that is, when it is in a collective
motion with velocity $\mathbf{V}$ relative to $K$. 

Now, applying all these concepts, what the RP states is the following:
\begin{equation}
T_{\mathbf{V}}\left(M_{\mathbf{V}}(F)\right)=P_{\mathbf{V}}(F)\,\,\,\,\mbox{ for all }F\in\mathcal{E}\label{eq:RP-math1}
\end{equation}
or equivalently, 
\begin{equation}
P_{\mathbf{V}}(F)\subset R'\mbox{ and }M_{\mathbf{V}}(F)=T_{\mathbf{V}}^{-1}\left(P_{\mathbf{V}}(F)\right)\,\,\,\,\mbox{ for all }F\in\mathcal{E}\label{eq:RP-math1a}
\end{equation}

\R \label{meg:(M)}The following is a minimal requirement for $M_{\mathbf{V}}:\,\mathcal{E}\rightarrow\mathcal{E}$
to have the assumed physical meaning; that is, a minimal requirement
for the RP to be a meaningful statement:
\begin{lyxlist}{00.00.0000}
\item [{(M)}] Relations $F\in\mathcal{E}$ must describe situations which\emph{
}can be meaningfully characterized as such in which the system as
a whole is at rest or in motion with some velocity relative to a frame
of reference. 
\end{lyxlist}
Notice that this requirement says nothing about whether and how the
fact that the system as a whole is at rest or in motion with some
velocity is reflected in the solutions $F\in\mathcal{E}$. It does
not even require that this fact can be expressed in terms of $\xi_{1},\xi_{2},\ldots\xi_{n}$.
It only requires that each $F\in\mathcal{E}$ belong to a physical
situation in which it is meaningful to say---perhaps in terms of quantities
different from $\xi_{1},\xi_{2},\ldots\xi_{n}$---that the system
is at rest or in motion relative to a reference frame. How a physical
situation can be characterized as such in which the system is at rest
or in motion is a separate problem, which will be discussed in section~\ref{sec:Is-relativity-principle}
in the particular case of a particles + electromagnetic field system.
\hfill{} $\lrcorner$

\R\label{Meg:kontingencia}Notice that, for a given fixed $F$, everything
on the right hand side of the equation in (\ref{eq:RP-math1a}), $P_{V}$
and $T_{V}$, are determined \emph{only} \emph{by} the physical behaviors
of \emph{the} \emph{measuring equipments} when they are in various
states of motion. In contrast, the meaning of the left hand side,
$M_{V}(F)$, depends on the physical behavior of \emph{the object
physical system} described by $F$ and $M_{V}(F)$, when it is in
various states of motion. That is to say, the two sides of the equation
reflect the behaviors\emph{ }of\emph{ different parts} of the physical
reality; and the RP expresses a law-like regularity between the behaviors\emph{
}of\emph{ }these different parts\emph{.}\hfill{} $\lrcorner$

\R\label{meg:example}Let us illustrate these concepts with a well-known
textbook example of a static versus uniformly moving charged particle.
The static field of a charge $q$ being at \emph{rest} at point $\mathbf{r}_{0}$
in $K$ is the following%
\footnote{In this example, $\mathbf{E},\mathbf{B}$ and $q$ denote the usual
textbook concepts of field strengths and charge, which are not entirely
the same as the ones we will introduce in the next section.%
}:
\begin{equation}
F\,\,\,\left\{ \begin{alignedat}{1}E_{x} & =\frac{q\left(x-x_{0}\right)}{\left(\left(x-x_{0}\right)^{2}+\left(y-y_{0}\right)^{2}+\left(z-z_{0}\right)^{2}\right)^{\nicefrac{3}{2}}}\\
E_{y} & =\frac{q\left(y-y_{0}\right)}{\left(\left(x-x_{0}\right)^{2}+\left(y-y_{0}\right)^{2}+\left(z-z_{0}\right)^{2}\right)^{\nicefrac{3}{2}}}\\
E_{z} & =\frac{q\left(z-z_{0}\right)}{\left(\left(x-x_{0}\right)^{2}+\left(y-y_{0}\right)^{2}+\left(z-z_{0}\right)^{2}\right)^{\nicefrac{3}{2}}}\\
B_{x} & =0\\
B_{y} & =0\\
B_{z} & =0
\end{alignedat}
\right.\label{eq:Coulomb field}
\end{equation}

The stationary field of a charge $q$ \emph{moving} at constant velocity
$\mathbf{V}=\left(V,0,0\right)$ relative to $K$ can be obtained
by solving the equations of ED (in $K$) with the time-depending source
(for example, Jackson 1999, pp. 661--665): 

\begin{equation}
M_{\mathbf{V}}(F)\,\,\,\left\{ \begin{alignedat}{1}E_{x} & =\frac{qX_{0}}{\left(X_{0}^{2}+\left(y-y_{0}\right)^{2}+\left(z-z_{0}\right)^{2}\right)^{\nicefrac{3}{2}}}\\
E_{y} & =\frac{\gamma q\left(y-y_{0}\right)}{\left(X_{0}^{2}+\left(y-y_{0}\right)^{2}+\left(z-z_{0}\right)^{2}\right)^{\nicefrac{3}{2}}}\\
E_{z} & =\frac{\gamma q\left(z-z_{0}\right)}{\left(X_{0}^{2}+\left(y-y_{0}\right)^{2}+\left(z-z_{0}\right)^{2}\right)^{\nicefrac{3}{2}}}\\
B_{x} & =0\\
B_{y} & =-c^{-2}VE_{z}\\
B_{z} & =c^{-2}VE_{y}
\end{alignedat}
\right.\label{eq:Coulomb-mozgo}
\end{equation}
where $X_{0}=\gamma\left(x-\left(x_{0}+Vt\right)\right)$. Note that
both solutions (\ref{eq:Coulomb field}) and (\ref{eq:Coulomb-mozgo})
satisfy condition (M), as it will be seen in section~\ref{sec:Is-relativity-principle}.

Now, we form the same expressions as (\ref{eq:Coulomb field}) but
in the \emph{primed} variables of the co-moving reference frame $K'$:
\begin{equation}
P_{\mathbf{V}}\left(F\right)\,\,\,\left\{ \begin{alignedat}{1}E'_{x} & =\frac{q'\left(x'-x'_{0}\right)}{\left(\left(x'-x'_{0}\right)^{2}+\left(y'-y'_{0}\right)^{2}+\left(z'-z'_{0}\right)^{2}\right)^{\nicefrac{3}{2}}}\\
E'_{y} & =\frac{q'\left(y'-y'_{0}\right)}{\left(\left(x'-x'_{0}\right)^{2}+\left(y'-y'_{0}\right)^{2}+\left(z'-z'_{0}\right)^{2}\right)^{\nicefrac{3}{2}}}\\
E'_{z} & =\frac{q'\left(z'-z'_{0}\right)}{\left(\left(x'-x'_{0}\right)^{2}+\left(y'-y'_{0}\right)^{2}+\left(z'-z'_{0}\right)^{2}\right)^{\nicefrac{3}{2}}}\\
B'_{x} & =0\\
B'_{y} & =0\\
B'_{z} & =0
\end{alignedat}
\right.\label{eq:Coulomb-vesszos}
\end{equation}
By means of the Lorentz transformation rules of the space-time coordinates,
the field strengths and the electric charge, one can express (\ref{eq:Coulomb-vesszos})
in terms of the original variables of $K$:

\begin{equation}
T_{\mathbf{V}}^{-1}\left(P_{\mathbf{V}}(F)\right)\,\,\,\left\{ \begin{alignedat}{1}E_{x} & =\frac{qX_{0}}{\left(X_{0}^{2}+\left(y-y_{0}\right)^{2}+\left(z-z_{0}\right)^{2}\right)^{\nicefrac{3}{2}}}\\
E_{y} & =\frac{\gamma q\left(y-y_{0}\right)}{\left(X_{0}^{2}+\left(y-y_{0}\right)^{2}+\left(z-z_{0}\right)^{2}\right)^{\nicefrac{3}{2}}}\\
E_{z} & =\frac{\gamma q\left(z-z_{0}\right)}{\left(X_{0}^{2}+\left(y-y_{0}\right)^{2}+\left(z-z_{0}\right)^{2}\right)^{\nicefrac{3}{2}}}\\
B_{x} & =0\\
B_{y} & =-c^{-2}VE_{z}\\
B_{z} & =c^{-2}VE_{y}
\end{alignedat}
\right.
\end{equation}
We find that the result is indeed the same as (\ref{eq:Coulomb-mozgo})
describing the field of the moving charge: $M_{\mathbf{V}}(F)=T_{\mathbf{V}}^{-1}\left(P_{\mathbf{V}}(F)\right)$.
That is to say, the RP seems to be true in this particular case. \hfill{}
$\lrcorner$

\medskip{}

\noindent Now we have a strict mathematical formulation of the RP
for a physical system described by a system of equations $\mathcal{E}$.
Remarkably, however, we still have not encountered the concept of
{}``covariance'' of equations $\mathcal{E}$. The reason is that
the RP and the covariance of equations $\mathcal{E}$ are not equivalent---in
contrast to what many believe. In fact, the logical relationship between
the two conditions is much more complex. To see this relationship
in more details, we previously need to clarify a few things. 

Consider the following two sets: $P_{\mathbf{V}}(\mathcal{E})=\{P_{\mathbf{V}}(F)|F\in\mathcal{E}\}$
and $T_{\mathbf{V}}(\mathcal{E})=\{T_{\mathbf{V}}(F)|F\in\mathcal{E}\}$.
Since a system of equations can be identified with its set of solutions,
$P_{\mathbf{V}}(\mathcal{E})\subset2^{\Omega'}$ and $T_{\mathbf{V}}(\mathcal{E})\subset2^{R'}$
can be regarded as two systems of equations for functional relations
between $\xi'_{1},\xi'_{2},\ldots\xi'_{n}$. In the primed variables,
$P_{\mathbf{V}}(\mathcal{E})$ has {}``the same form'' as $\mathcal{E}$.
Nevertheless, it can be the case that $P_{\mathbf{V}}(\mathcal{E})$
does not express a true physical law, in the sense that its solutions
do not necessarily describe true physical situations. In contrast,
$T_{\mathbf{V}}(\mathcal{E})$ is nothing but $\mathcal{E}$ expressed
in variables\emph{ $\xi'_{1},\xi'_{2},\ldots\xi'_{n}$}. 

Now, covariance intuitively means that equations $\mathcal{E}$ {}``preserve
their forms against the transformation $T_{\mathbf{V}}$''. That
is, in terms of the formalism we developed: 
\begin{equation}
T_{\mathbf{V}}(\mathcal{E})=P_{\mathbf{V}}(\mathcal{E})\label{eq:kovi}
\end{equation}
or, equivalently,
\begin{equation}
P_{\mathbf{V}}(\mathcal{E})\subset2^{R'}\,\mbox{ and }\,\mathcal{E}=T_{\mathbf{V}}^{-1}\left(P_{\mathbf{V}}(\mathcal{E})\right)\label{eq:kovi-a}
\end{equation}

The first thing we have to make clear is that---even if we know or
presume that it holds---covariance (\ref{eq:kovi-a}) is obviously
\emph{not sufficient} for the RP (\ref{eq:RP-math1a}). For, (\ref{eq:kovi-a})
only guarantees the invariance of the set of solutions, $\mathcal{E}$,
against $T_{\mathbf{V}}^{-1}\circ P_{\mathbf{V}}$ , but it says nothing
about which solution of $\mathcal{E}$ corresponds to which solution.
In Bell's words:
\begin{quote}
Lorentz invariance alone shows that for any state of a system at rest
there is a corresponding `primed' state of that system in motion.
But it does not tell us that if the system is set anyhow in motion,
it will actually go into the 'primed' of the original state, rather
than into the `prime' of some \emph{other} state of the original system.
(Bell 1987, p.~75)
\end{quote}
While it is the very essence of the RP that the solution $M_{\mathbf{V}}(F)$,
describing the system \emph{in motion} relative to $K$, corresponds
to solution $T_{\mathbf{V}}^{-1}\circ P_{\mathbf{V}}(F)$.%
\footnote{The difference between covariance and the RP is obvious from the well-known
applications of the RP. For example, what we use in the derivation
of electromagnetic field of a uniformly moving point charge (Remark~\ref{meg:3a})
is not the covariance of the equations, but statement (\ref{eq:RP-math1a}),
that is, what the RP claims about the solutions of the equations in
details. %
}

In a precise sense, covariance is not only not sufficient for the
RP, but it is \emph{not even necessary} 
\begin{figure}
\begin{centering}
\includegraphics[width=0.8\textwidth]{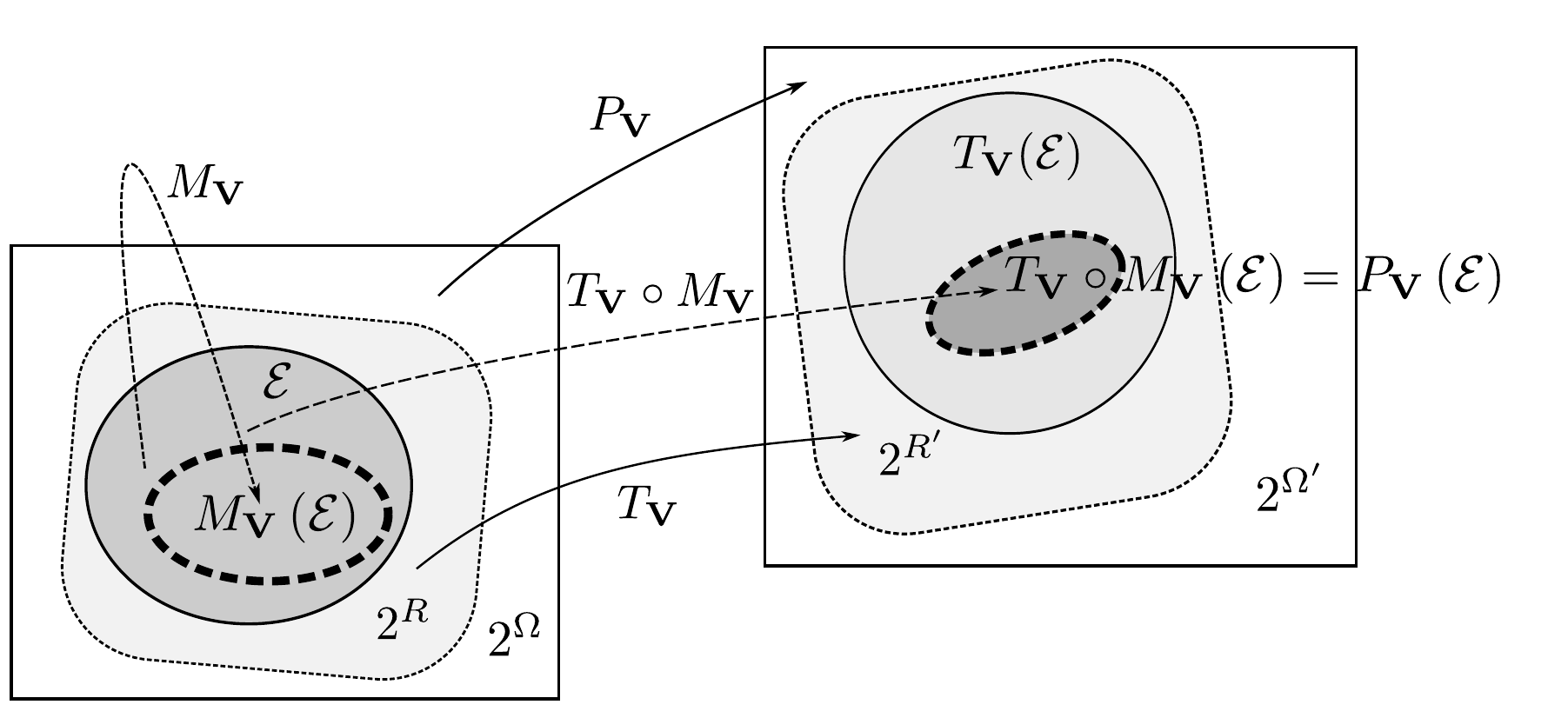}
\par\end{centering}

\caption{The RP only implies that $T_{\mathbf{V}}(\mathcal{E})\supseteq T_{\mathbf{V}}\circ M_{\mathbf{V}}\left(\mathcal{E}\right)=P_{\mathbf{V}}\left(\mathcal{E}\right)$.
Covariance of $\mathcal{E}$ would require that $T_{\mathbf{V}}(\mathcal{E})=P_{\mathbf{V}}(\mathcal{E})$,
which is generally not the case\label{fig:Relativity-principle-only}}
\end{figure}
 (Fig.~\ref{fig:Relativity-principle-only}). The RP only implies
that 
\begin{equation}
T_{\mathbf{V}}(\mathcal{E})\supseteq T_{\mathbf{V}}\left(M_{\mathbf{V}}\left(\mathcal{E}\right)\right)=P_{\mathbf{V}}\left(\mathcal{E}\right)
\end{equation}
 (\ref{eq:RP-math1}) implies (\ref{eq:kovi}) only if we have the
following extra condition: 
\begin{equation}
M_{\mathbf{V}}\left(\mathcal{E}\right)=\mathcal{E}\label{eq:RE-kov2}
\end{equation}

We will return to the problem of how little we can say about $M_{\mathbf{V}}$
and condition (M) in general; what we have to see here is that the
RP implies covariance only in conjunction with (\ref{eq:RE-kov2}).\emph{ }

What is the situation in ED? 
\begin{itemize}
\item As we will see in section~\ref{sec:Is-relativity-principle}, the
very concept of $M{}_{\mathbf{V}}$ is problematic in ED. Consequently,
there is no guarantee that condition \eqref{eq:RE-kov2} is satisfied. 
\item In any event, we will show the covariance of the Maxwell--Lorentz
equations, \emph{independently of the RP}; in the sense that we will
determine the transformation of the electrodynamical quantities, independently
of the RP---and without presuming the covariance, of course---and
will see that the equations are covariant against these transformations.
\item As we have seen, the covariance of the equations is not sufficient;
the question whether the RP holds in ED will be discussed in section~\ref{sec:Is-relativity-principle}.
\end{itemize}
\R\label{meg:1}The notion of $M_{\mathbf{V}}$ plays a crucial role.
Formally, one could say, the RP is \emph{relative} to the definition
of $M_{\mathbf{V}}$. Therefore, the physical content of the RP depends
on how this concept is physically understood. But, what does it mean
to say that a physical system is the same and of the same behavior
as the one described by $F$, except that it is, as a whole, in a
collective motion with velocity $\mathbf{V}$ relative to $K$? And,
first of all, what does it mean to say that a physical system, as
a whole, is at rest or in collective motion with some velocity relative
to a reference frame? Without answering this crucial questions the
RP is meaningless. 

In fact, the same questions can be asked with respect to the definitions
of quantities $\xi'_{1},\xi'_{2},\ldots\xi'_{n}$---and, therefore,
with respect to the meanings of $T_{\mathbf{V}}$ and $P_{\mathbf{V}}$.
For, $\xi'_{1},\xi'_{2},\ldots\xi'_{n}$ are not simply arbitrary
variables assigned to reference frame $K'$, in one-to-one relations
with $\xi_{1},\xi_{2},\ldots\xi_{n}$, but the physical quantities
obtainable by means of the same operations with the same measuring
equipments as in the operational definitions of $\xi_{1},\xi_{2},\ldots\xi_{n}$,
except that everything is in a collective motion with velocity $\mathbf{V}$.
Therefore, we should know what we mean by {}``the same measuring
equipment but in collective motion''. From this point of view, it
does not matter whether the system in question is the object to be
observed or a measuring equipment involved in the observation. 

One might claim that $M_{\mathbf{V}}(F)$, describing the moving system,
is equal to the {}``Lorentz boosted solution'' \emph{by definition}:
\begin{eqnarray}
M_{\mathbf{V}}(F) & \overset{^{def}}{=} & T_{\mathbf{V}}^{-1}\left(P_{\mathbf{V}}(F)\right)\label{eq:haegyenlok}
\end{eqnarray}
 But a little reflection shows that it is, in fact, untenable:
\begin{lyxlist}{00.00.0000}
\item [{(a)}] In this case, (\ref{eq:RP-math1a}) would read 
\begin{equation}
T_{\mathbf{V}}^{-1}\left(P_{\mathbf{V}}(F)\right)=T_{\mathbf{V}}^{-1}\left(P_{\mathbf{V}}(F)\right)\label{eq:RP-tautologi}
\end{equation}
That is, the RP would become a tautology; a statement which is always
true, independently of any contingent fact of nature; independently
of the actual behavior of moving physical objects; and independently
of the actual empirical meanings of physical quantities $\xi'_{1},\xi'_{2},\ldots\xi'_{n}$.
But, the RP is supposed to be a fundamental \emph{law of nature}.
Note that a tautology is entirely different from a fundamental principle,
even if the principle is used as a fundamental hypothesis or fundamental
premise of a theory, from which one derives further physical statements.
For, a fundamental premise, as expressing a contingent fact of nature,
is potentially falsifiable by testing its consequences; a tautology
is not.
\item [{(b)}] Even if accepted, (\ref{eq:haegyenlok}) can provide physical
meaning to $M_{\mathbf{V}}(F)$ only if we know the meanings of $T_{\mathbf{V}}$
and $P_{\mathbf{V}}$, that is, if we know the empirical meanings
of the quantities denoted by $\xi'_{1},\xi'_{2},\ldots\xi'_{n}$.
But, the physical meaning of $\xi'_{1},\xi'_{2},\ldots\xi'_{n}$ are
obtained from the operational definitions: they are the quantities
obtained by {}``the same measurements with the same equipments when
they are, as a whole, co-moving with\emph{ $K'$ }with velocity $\mathbf{V}$
relative to $K$''. Symbolically, we need, priory, the concepts of
$M_{\mathbf{V}}(\xi_{i}\mbox{-}equipment\, at\, rest)$. And this
is a conceptual circularity: in order to have the concept of what
it is to be an $M_{\mathbf{V}}(brick\, at\, rest)$ the (size)' of
which we would like to ascertain, we need to have the concept of what
it is to be an $M_{\mathbf{V}}(measuring\, rod\, at\, rest)$---which
is exactly the same conceptual problem. 
\item [{(c)}] One might claim that we do not need to specify the concepts
of $M_{\mathbf{V}}(\xi_{i}\mbox{-}equipment\, at\, rest)$ in order
to know the \emph{values} of quantities $\xi'_{1},\xi'_{2},\ldots\xi'_{n}$
we obtain by the measurements with the moving equipments, given that
we can know the transformation rule $T_{\mathbf{V}}$ independently
of knowing the operational definitions of $\xi'_{1},\xi'_{2},\ldots\xi'_{n}$.
Typically, $T_{\mathbf{V}}$ is thought to be derived from the assumption
that the RP \eqref{eq:RP-math1a} holds. If however $M_{\mathbf{V}}$
is, by definition, equal to $T_{\mathbf{V}}^{-1}\circ P_{\mathbf{V}}$,
then in place of \eqref{eq:RP-math1a} we have the tautology (\ref{eq:RP-tautologi}),
which does not determine $T_{\mathbf{V}}$.
\item [{(d)}] Therefore, unsurprisingly, it is not the RP from which transformation
rule $T_{\mathbf{V}}$ is routinely deduced, but the covariance \eqref{eq:kovi-a}.
As we have seen, however, covariance is, in general, neither sufficient
nor necessary for the RP. Whether \eqref{eq:RP-math1a} implies \eqref{eq:kovi-a}
hinges on the physical fact whether \eqref{eq:RE-kov2} is satisfied.
But, if $M_{\mathbf{V}}$ is taken to be $T_{\mathbf{V}}^{-1}\circ P_{\mathbf{V}}$
by definition, the RP becomes true---in the form of tautology (\ref{eq:RP-tautologi})---but
does not imply covariance $T_{\mathbf{V}}^{-1}\circ P_{\mathbf{V}}(\mathcal{E})=\mathcal{E}$.
\item [{(e)}] Even if we assume that a {}``transformation rule'' function
$\phi'\circ T_{\mathbf{V}}\circ\phi^{-1}$ were derived from some
independent premises---from the independent assumption of covariance,
for example---how do we know that the $T_{\mathbf{V}}$ we obtained
and the quantities of values $\phi'\circ T_{\mathbf{V}}\circ\phi^{-1}\left(\xi_{1},\xi_{2},\ldots\xi_{n}\right)$
are correct plugins for the RP? How could we verify that $\phi'\circ T_{\mathbf{V}}\circ\phi^{-1}\left(\xi_{1},\xi_{2},\ldots\xi_{n}\right)$
are indeed the values measured by a moving observer applying the same
operations with the same measuring equipments, etc.?---without having
an independent concept of $M_{\mathbf{V}}$, at least for the measuring
equipments?
\item [{(f)}] One could argue that we do not need such a verification;
$\phi'\circ T_{\mathbf{V}}\circ\phi^{-1}\left(\xi_{1},\xi_{2},\ldots\xi_{n}\right)$
can be regarded \emph{as the empirical definition} of the primed quantities:
\begin{equation}
\left(\xi'_{1},\xi'_{2},\ldots\xi'_{n}\right)\overset{^{def}}{=}\phi'\circ T_{\mathbf{V}}\circ\phi^{-1}\left(\xi_{1},\xi_{2},\ldots\xi_{n}\right)\label{eq:trafo-mint-def}
\end{equation}
This is of course logically possible. The operational definition of
the primed quantities would say: ask the observer at rest in $K$
to measure $\xi_{1},\xi_{2},\ldots\xi_{n}$ with the measuring equipments
at rest in $K$, and then perform the mathematical operation \eqref{eq:trafo-mint-def}.
In this way, however, even the transformation rules would become tautologies;
they would be true, no matter how the things are in the physical world.
\item [{(g)}] Someone might claim that the identity of $M_{\mathbf{V}}$
with $T_{\mathbf{V}}^{-1}\circ P_{\mathbf{V}}$ is not a simple stipulation
but rather an analytic truth which follows from the identity of the
two \emph{concepts}. Still, if that were the case, RP would be a statement
which is true in all possible worlds; independently of any contingent
fact of nature; independently of the actual behavior of moving physical
objects.
\item [{(h)}] On the contrary, as we have already pointed out in Remark~\ref{Meg:kontingencia},
$M_{\mathbf{V}}(F)$ and $T_{\mathbf{V}}^{-1}\left(P_{\mathbf{V}}(F)\right)$
are \emph{different concepts}, referring to different features of
different parts of the physical reality. Any connection between the
two things must be a contingent fact of the world. 
\item [{(i)}] $T_{\mathbf{V}}^{-1}\circ P_{\mathbf{V}}$ is a $2^{R}\rightarrow2^{R}$
map which is completely determined\emph{ }by\emph{ }the physical behaviors
of the\emph{ }measuring equipments. On the other hand, whether the
elements of $\mathcal{E}\subset2^{R}$ satisfy condition (M) and whether
$T_{\mathbf{V}}^{-1}\circ P_{\mathbf{V}}$ preserves $\mathcal{E}$
depend on the actual physical properties of the object physical system.\hfill{}
$\lrcorner$\medskip{}

\end{lyxlist}
\R\label{meg:kovariancia-uj}Finally, let us note a few important
facts which can easily be seen in the formalism we developed:
\begin{lyxlist}{00.00.0000}
\item [{(a)}] The covariance of a set of equations $\mathcal{E}$ does
\emph{not} imply the covariance of a subset of equations separately.
It is because a smaller set of equations corresponds to an $\mathcal{E}^{*}\subset2^{R}$
such that $\mathcal{E}\subset\mathcal{E}^{*}$; and it does not follow
from (\ref{eq:kovi}) that $T_{\mathbf{V}}(\mathcal{E}^{*})=P_{\mathbf{V}}(\mathcal{E}^{*})$.
\item [{(b)}] Similarly, the covariance of a set of equations $\mathcal{E}$
does \emph{not} guarantee the covariance of an arbitrary set of equations
which is only satisfactory to $\mathcal{E}$; for example, when the
solutions of $\mathcal{E}$ are restricted by some extra conditions.
Because from (\ref{eq:kovi}) it does not follow that $T_{\mathbf{V}}(\mathcal{E}^{*})=P_{\mathbf{V}}(\mathcal{E}^{*})$
for an arbitrary $\mathcal{E}^{*}\subset\mathcal{E}$.
\item [{(c)}] The same holds, of course, for the combination of cases (a)
and (b); for example, when we have a smaller set of equations $\mathcal{E}^{*}\supset\mathcal{E}$
together with some extra conditions $\psi\subset2^{R}$. For, (\ref{eq:kovi})
does not imply that $T_{\mathbf{V}}(\mathcal{E}^{*}\cap\psi)=P_{\mathbf{V}}(\mathcal{E}^{*}\cap\psi)$.
\item [{(d)}] However, covariance is guaranteed if a covariant set of equations
is restricted with a \emph{covariant} set of extra conditions; because
$T_{\mathbf{V}}(\mathcal{E})=P_{\mathbf{V}}(\mathcal{E})$ and $T_{\mathbf{V}}(\psi)=P_{\mathbf{V}}(\psi)$
trivially imply that $T_{\mathbf{V}}(\mathcal{E}\cap\psi)=P_{\mathbf{V}}(\mathcal{E}\cap\psi)$.\hfill{}
$\lrcorner$\medskip{}

\end{lyxlist}

\section{Operational definitions of electrodynamical quantities in $K$}

Now we turn to the operational definitions of the fundamental electrodynamical
quantities in a single reference frame $K$ and to the basic observational
facts about these quantities. 

The operational definition of a physical quantity requires the specification
of \emph{etalon} physical objects and standard physical processes
by means of which the value of the quantity is ascertained. In case
of electrodynamical quantities the only {}``device'' we need is
a point-like test particle, and the standard measuring procedures
by which the kinematical properties of the test particle are ascertained. 

So, assume we have chosen an \emph{etalon} test particle, and let
$\mathbf{r}^{etalon}(t)$, $\mathbf{v}^{etalon}(t)$, $\mathbf{a}^{etalon}(t)$
denote its position, velocity and acceleration at time $t$. It is
assumed that we are able to set the \emph{etalon} test particle into
motion with arbitrary velocity $\mathbf{v}^{etalon}<c$ at arbitrary
location. We will need more {}``copies'' of the \emph{etalon} test
particle:

\paragraph*{Definition~(D0)}

A particle $e$ is called \emph{test particle} if for all $\mathbf{r}$
and $t$
\begin{equation}
\mathbf{v}^{e}\left(t\right)\biggl|_{\mathbf{r}^{e}\left(t\right)=\mathbf{r}}=\mathbf{v}^{etalon}\left(t\right)\biggl|_{\mathbf{r}^{etalon}\left(t\right)=\mathbf{r}}
\end{equation}
implies
\begin{equation}
\mathbf{a}^{e}\left(t\right)\biggl|_{\mathbf{r}^{e}\left(t\right)=\mathbf{r}}=\mathbf{a}^{etalon}\left(t\right)\biggl|_{\mathbf{r}^{etalon}\left(t\right)=\mathbf{r}}
\end{equation}
(The {}``restriction signs'' refer to \emph{physical} situations;
for example, $|_{\mathbf{r}^{e}\left(t\right)=\mathbf{r}}$ indicates
that the test particle $e$ is at point $\mathbf{r}$ at time $t$.)
\medskip{}

\noindent Note, that some of the definitions and statements below
require the existence of many test particles; which is, of course,
a matter of empirical fact, and will be provided by (E0) below.

First we define the electric and magnetic field strengths. The only
measuring device we need is a test particle being at rest relative
to $K$.

\paragraph*{Definition~(D1)}

\noindent \emph{Electric field strength} at point $\mathbf{r}$ and
time $t$ is defined as the acceleration of an arbitrary test particle
$e$, such that $\mathbf{r}^{e}(t)=\mathbf{r}$ and $\mathbf{v}^{e}(t)=0$:
\begin{equation}
\mathbf{E}\left(\mathbf{r},t\right)\overset{^{def}}{=}\left.\mathbf{a}^{e}(t)\right|_{\mathbf{r}^{e}(t)=\mathbf{r};\,\mathbf{v}^{e}(t)=0}\label{eq:E-def}
\end{equation}
\medskip{}

\noindent Magnetic field strength is defined by means of how the acceleration
$\mathbf{a}^{e}$ of the rest test particle changes with an infinitesimal
perturbation of its state of rest, that is, if an infinitesimally
small velocity $\mathbf{v}^{e}$ is imparted to the particle. Of course,
we cannot perform various small perturbations simultaneously on one
and the same rest test particle, therefore we perform the measurements
on many rest test particles with various small perturbations. Let
$\delta\subset\mathbb{R}^{3}$ be an arbitrary infinitesimal neighborhood
of $0\in\mathbb{R}^{3}$. First we define the following function:
\begin{eqnarray}
\mathbf{U}^{\mathbf{r},t} & : & \mathbb{R}^{3}\supset\delta\rightarrow\mathbb{R}^{3}\nonumber \\
 &  & \mathbf{U}^{\mathbf{r},t}(\mathbf{v})\overset{^{def}}{=}\left.\mathbf{a}^{e}(t)\right|_{\mathbf{r}^{e}(t)=\mathbf{r};\,\mathbf{v}^{e}(t)=\mathbf{v}}\label{eq:U-def}
\end{eqnarray}
Obviously, $\mathbf{U}^{\mathbf{r},t}(0)=\mathbf{E}\left(\mathbf{r},t\right)$.

\paragraph*{Definition~(D2)}

\noindent \emph{Magnetic field strength} at point $\mathbf{r}$ and
time $t$ is
\begin{equation}
\mathbf{B}(\mathbf{r},t)\overset{^{def}}{=}\left.\left(\begin{array}{c}
\partial_{v_{z}}U_{y}^{\mathbf{r},t}\\
\partial_{v_{x}}U_{z}^{\mathbf{r},t}\\
\partial_{v_{y}}U_{x}^{\mathbf{r},t}
\end{array}\right)\right|_{\mathbf{v}=0}\label{eq:B-def}
\end{equation}

\noindent \medskip{}
Practically it means that one can determine the value of $\mathbf{B}(\mathbf{r},t)$,
with arbitrary precision, by means of measuring the accelerations
of a few test particles of velocity $\mathbf{v}^{e}\in\delta$.

Next we introduce the concepts of source densities:

\paragraph*{Definition~(D3)}

\begin{eqnarray}
\varrho\left(\mathbf{r},t\right) & \overset{^{def}}{=} & \nabla\cdot\mathbf{E}\left(\mathbf{r},t\right)\label{eq:ME1}\\
\mathbf{j}\left(\mathbf{r},t\right) & \overset{^{def}}{=} & c^{2}\nabla\times\mathbf{B}\left(\mathbf{r},t\right)-\partial_{t}\mathbf{E}\left(\mathbf{r},t\right)\label{eq:ME2}
\end{eqnarray}
are called \emph{active electric charge density} and \emph{active
electric current density,} respectively. \medskip{}

\noindent A simple consequence of the \emph{definitions} is that a
continuity equation holds for $\varrho$ and $\mathbf{j}$:
\begin{thm}
~
\begin{equation}
\partial_{t}\varrho\left(\mathbf{r},t\right)+\nabla\cdot\mathbf{j}\left(\mathbf{r},t\right)=0\label{eq:kontinuitas}
\end{equation}

\end{thm}
\R\label{meg:2}In our construction, the two Maxwell equations (\ref{eq:ME1})--(\ref{eq:ME2}),
are mere \emph{definitions} of the concepts of active electric charge
density and\emph{ }active electric current density. They do not contain
information whatsoever about how {}``matter produces electromagnetic
field''. And it is not because $\varrho\left(\mathbf{r},t\right)$
and $\mathbf{j}\left(\mathbf{r},t\right)$ are, of course, {}``unspecified
distributions'' in these {}``general laws'', but because $\varrho\left(\mathbf{r},t\right)$
and $\mathbf{j}\left(\mathbf{r},t\right)$ cannot be specified prior
to or at least independently of the field strengths $\mathbf{E}(\mathbf{r},t)$
and $\mathbf{B}(\mathbf{r},t)$. Again, because $\varrho\left(\mathbf{r},t\right)$
and $\mathbf{j}\left(\mathbf{r},t\right)$ are just abbreviations,
standing for the expressions on the right hand sides of (\ref{eq:ME1})--(\ref{eq:ME2}).
In other words, any statement about the {}``charge distribution''
will be a statement about $\nabla\cdot\mathbf{E}$, and any statement
about the {}``current distribution'' will be a statement about $c^{2}\nabla\times\mathbf{B}-\partial_{t}\mathbf{E}$.

The minimal claim is that this is a possible coherent construction.
Though we must add: equations (\ref{eq:ME1})--(\ref{eq:ME2}) could
be seen as contingent physical laws about the relationship between
the charge and current distributions and the electromagnetic field,
only if we had an independent empirical definition of charge. However,
we do not see how such a definition is possible, without encountering
circularities. (Also see Remark~\ref{meg:3}.) \hfill{} $\lrcorner$\medskip{}

\noindent The operational definitions of the field strengths and the
source densities are based on the kinematical properties of the test
particles. The following definition describes the concept of a charged
point-like particle, in general.

\paragraph*{Definition (D4)}

A particle $b$ is called \emph{charged point-particle} of \emph{specific
passive electric charge} $\pi^{b}$ and of \emph{active electric charge}
$\alpha^{b}$ if the following is true:
\begin{enumerate}
\item It satisfies the relativistic Lorentz equation,

\noindent 
\begin{eqnarray}
\gamma\left(\mathbf{v}^{b}\left(t\right)\right)\mathbf{a}^{b}(t) & = & \pi^{b}\left\{ \mathbf{E}\left(\mathbf{r}^{b}\left(t\right),t\right)+\mathbf{v}^{b}\left(t\right)\times\mathbf{B}\left(\mathbf{r}^{b}\left(t\right),t\right)\right.\nonumber \\
 &  & \left.-c^{-2}\mathbf{v}^{b}\left(t\right)\left(\mathbf{v}^{b}\left(t\right)\mathbf{\cdot E}\left(\mathbf{r}^{b}\left(t\right),t\right)\right)\right\} \label{eq:E1'}
\end{eqnarray}

\item \noindent If it is the only particle whose worldline intersects a
given space-time region $\Omega$, then for all $(\mathbf{r},t)\in\Omega$
the source densities are of the following form: 
\begin{eqnarray}
\varrho\left(\mathbf{r},t\right) & = & \alpha^{b}\delta\left(\mathbf{r}-\mathbf{r}^{b}\left(t\right)\right)\label{eq:(E3)1}\\
\mathbf{j}\left(\mathbf{r},t\right) & = & \alpha^{b}\delta\left(\mathbf{r}-\mathbf{r}^{b}\left(t\right)\right)\mathbf{v}^{b}\left(t\right)\label{eq:(E3)2}
\end{eqnarray}

\end{enumerate}
where $\mathbf{r}^{b}\left(t\right)$, $\mathbf{v}^{b}\left(t\right)$
and $\mathbf{a}^{b}\left(t\right)$ are the particle's position, velocity
and acceleration. The ratio $\mu^{b}\overset{^{def}}{=}\alpha^{b}/\pi^{b}$
is called the \emph{electric inertial rest mass} of the particle.

\R\label{meg:3} Of course, \eqref{eq:E1'}is equivalent to the standard
form of the Lorentz equation: 
\begin{equation}
\frac{d}{dt}\left(\gamma\left(\mathbf{v}\left(t\right)\right)\mathbf{v}\left(t\right)\right)=\pi\left\{ \mathbf{E}\left(\mathbf{r}\left(t\right),t\right)+\mathbf{v}\left(t\right)\times\mathbf{B}\left(\mathbf{r}\left(t\right),t\right)\right\} 
\end{equation}
with $\pi=q/m$ in the usual terminology, where $q$ is the passive
electric charge and $m$ is the inertial (rest) mass of the particle---that
is why we call $\pi$ \emph{specific} passive electric charge. Nevertheless,
it must be clear that for all charged point-particles we introduced
\emph{two independent}, empirically meaningful and experimentally
testable quantities: specific passive electric charge $\pi$ and active
electric charge $\alpha$. There is no universal law-like relationship
between these two quantities: the ratio between them varies from particle
to p1article. In the traditional sense, this ratio is, however, nothing
but the particle's rest mass.

We must emphasize that the concept of mass so obtained, as defined
by only means of electrodynamical quantities, is essentially related
to ED, that is to say, to electromagnetic interaction. There seems
no way to give a consistent and non-circular operational definition
of inertial mass in general, independently of the context of a particular
type of physical interaction. Without entering here into the detailed
discussion of the problem, we only mention that, for example, Weyl's
commonly accepted definition (Jammer 2000, pp. 8--10) and all similar
definitions based on the conservation of momentum in particle collisions
suffer from the following difficulty. There is no {}``collision''
as a purely {}``mechanical'' process. During a collision the particles
are moving in a physical field---or fields---of interaction. Therefore:
1)~the system of particles, separately, cannot be regarded as a closed
system;\emph{ }2)~the inertial properties of the particles, in fact,
reveal themselves in the interactions with the field.\emph{ }Thus,
the concepts of inertial rest mass belonging to different interactions
differ from each other; whether they are equal (proportional) to each
other is a matter of contingent fact of nature.\emph{ }\hfill{} $\lrcorner$\medskip{}

\R\label{meg:4}The choice of the \emph{etalon} test particle is,
of course, a matter of convention, just as the definitions (D0)--(D4)
themselves. It is important to note that all these conventional factors
play a constitutive role in the fundamental concepts of ED (Reichenbach
1965). With these choices we not only make semantic\emph{ }conventions
determining the meanings of the terms, but also make a decision about
the body of concepts by means of which we grasp physical reality.\emph{
}There are a few things, however, that must be pointed out:
\begin{lyxlist}{00.00.0000}
\item [{(a)}] This kind of conventionality does not mean that the physical
quantities defined in (D0)--(D4) cannot describe \emph{objective}
features of physical reality. It only means that we make a decision
which objective features of reality we are dealing with. With another
body of conventions we have another body of physical concepts/physical
quantities and another body of empirical facts.
\item [{(b)}] \noindent On the other hand, it does not mean either that
our knowledge of the physical world would not be objective but a product
of our conventions. If two theories obtained by starting with two
different bodies of conventions are complete enough accounts of the
physical phenomena, then they describe the same reality, expressed
in terms of different physical quantities. Let us spell out an example:
Definition \eqref{eq:ME2} is entirely conventional---no objective
fact of the world determines the formula on the right hand side. Therefore,
we could make another choice, say,
\begin{equation}
\mathbf{j}_{\Theta}\left(\mathbf{r},t\right)\overset{^{def}}{=}\Theta^{2}\nabla\times\mathbf{B}\left(\mathbf{r},t\right)-\partial_{t}\mathbf{E}\left(\mathbf{r},t\right)\label{eq:tetasje}
\end{equation}
with some $\Theta\neq c$. At first sight, one might think that this
choice will alter the speed of electromagnetic waves. This is however
not the case. It will be an empirical fact \emph{about} $\mathbf{j}_{\Theta}\left(\mathbf{r},t\right)$
that if a particle $b$ is the only one whose worldline intersects
a given space-time region $\Omega$, then for all $(\mathbf{r},t)\in\Omega$
\begin{eqnarray}
\mathbf{j}_{\Theta}\left(\mathbf{r},t\right) & = & \alpha^{b}\delta\left(\mathbf{r}-\mathbf{r}^{b}\left(t\right)\right)\mathbf{v}^{b}\left(t\right)\nonumber \\
 &  & +\left(\Theta^{2}-c^{2}\right)\nabla\times\mathbf{B}(\mathbf{r},t)\label{eq:jeteta}
\end{eqnarray}
Now, consider a region where there is no particle. Taking into account
\eqref{eq:jeteta}, we have \eqref{eq:ME3}--\eqref{eq:ME4} and 
\begin{eqnarray}
\nabla\cdot\mathbf{E}(\mathbf{r},t) & = & 0\\
\Theta^{2}\nabla\times\mathbf{B}\left(\mathbf{r},t\right)-\partial_{t}\mathbf{E}\left(\mathbf{r},t\right) & = & \left(\Theta^{2}-c^{2}\right)\nabla\times\mathbf{B}(\mathbf{r},t)
\end{eqnarray}
which lead to the usual wave equation with propagation speed $c$.
(Of course, in this particular example, one of the possible choices,
namely $\Theta=c$, is distinguished by its simplicity. Note, however,
that simplicity is not an epistemologically unproblematic notion.)\hfill{}
$\lrcorner$
\end{lyxlist}

\section{Empirical facts of electrodynamics}

Both {}``empirical'' and {}``fact'' are used in different senses.
Statements (E0)--(E4) below are universal generalizations, rather
than statements of particular observations. Nevertheless we call them
{}``empirical facts'', by which we simply mean that they are truths
which can be acquired by \emph{a posteriori} means. Normally, they
can be considered as laws obtained by inductive generalization; statements
the truths of which can be, in principle, confirmed empirically. 

On the other hand, in the context of the consistency questions (Q3)
and (Q4), it is not important how these statements are empirically
confirmed. (E0)--(E4) can be regarded as axioms of the Maxwell--Lorentz
theory in $K$. What is important for us is that from these \emph{axioms},
in conjunction with the theoretical representations of the measurement
operations, there follow assertions about what the moving observer
in $K'$ observes. Section~\ref{sec:Observations-of-moving} will
be concerned with these consequences.

\paragraph*{(E0)}

There exist many enough test particles and we can settle them into
all required positions and velocities.

\medskip{}

\noindent Consequently, (D1)--(D4) are sound definitions. From observations
about $\mathbf{E}$, $\mathbf{B}$ and the charged point-particles,
we have further empirical facts:

\paragraph*{(E1)}

In all situations, the electric and magnetic field strengths satisfy
the following two Maxwell equations:

\noindent 
\begin{eqnarray}
\nabla\cdot\mathbf{B}\left(\mathbf{r},t\right) & = & 0\label{eq:ME3}\\
\nabla\times\mathbf{E}\left(\mathbf{r},t\right)+\partial_{t}\mathbf{B}\left(\mathbf{r},t\right) & = & 0\label{eq:ME4}
\end{eqnarray}

\paragraph*{(E2)}

Each particle is a charged point-particle, satisfying (D4) with some
specific passive electric charge $\pi$ and active electric charge
$\alpha$. This is also true for the test particles, with---as follows
from the definitions---specific passive electric charge $\pi=1$.

\paragraph*{(E3)}

If $b_{1}$, $b_{2}$,..., $b_{n}$ are the only particles whose worldlines
intersect a given space-time region $\Lambda$, then for all $(\mathbf{r},t)\in\Lambda$
the source densities are:
\begin{eqnarray}
\varrho\left(\mathbf{r},t\right) & = & \sum_{i=1}^{n}\alpha^{b_{i}}\delta\left(\mathbf{r}-\mathbf{r}^{b_{i}}\left(t\right)\right)\label{eq:add1}\\
\mathbf{j}\left(\mathbf{r},t\right) & = & \sum_{i=1}^{n}\alpha^{b_{i}}\delta\left(\mathbf{r}-\mathbf{r}^{b_{i}}\left(t\right)\right)\mathbf{v}^{b_{i}}\left(t\right)\label{eq:add2}
\end{eqnarray}
\medskip{}

Putting facts (E1)--(E3) together, we have the coupled Maxwell--Lorentz
equations:
\begin{eqnarray}
\nabla\cdot\mathbf{E}\left(\mathbf{r},t\right) & = & \sum_{i=1}^{n}\alpha^{b_{i}}\delta\left(\mathbf{r}-\mathbf{r}^{b_{i}}\left(t\right)\right)\label{eq:MLE1}\\
c^{2}\nabla\times\mathbf{B}\left(\mathbf{r},t\right)-\partial_{t}\mathbf{E}\left(\mathbf{r},t\right) & = & \sum_{i=1}^{n}\alpha^{b_{i}}\delta\left(\mathbf{r}-\mathbf{r}^{b_{i}}\left(t\right)\right)\mathbf{v}^{b_{i}}\left(t\right)\label{eq:MLE2}\\
\nabla\cdot\mathbf{B}\left(\mathbf{r},t\right) & = & 0\label{eq:MLE3}\\
\nabla\times\mathbf{E}\left(\mathbf{r},t\right)+\partial_{t}\mathbf{B}\left(\mathbf{r},t\right) & = & 0\label{eq:MLE4}\\
\gamma\left(\mathbf{v}^{b_{i}}\left(t\right)\right)\mathbf{a}^{b_{i}}(t) & = & \pi^{b_{i}}\left\{ \mathbf{E}\left(\mathbf{r}^{b_{i}}\left(t\right),t\right)+\mathbf{v}^{b_{i}}\left(t\right)\times\mathbf{B}\left(\mathbf{r}^{b_{i}}\left(t\right),t\right)\right.\,\,\,\nonumber \\
 &  & \left.-c^{-2}\mathbf{v}^{b_{i}}\left(t\right)\left(\mathbf{v}^{b_{i}}\left(t\right)\mathbf{\cdot E}\left(\mathbf{r}^{b_{i}}\left(t\right),t\right)\right)\right\} \label{eq:MLE5}\\
 &  & \,\,\,\,\,\,\,\,\,\,\,\,\,\,\,\,\,\,\,\,\,\,\,\,\,\,\,\,\,(i=1,2,\ldots n)\nonumber 
\end{eqnarray}

\noindent These are the fundamental equations of ED, describing an
interacting system of $n$ particles and the electromagnetic field. 

\R\label{meg:3a} Without entering into the details of the problem
of classical charged particles (Frisch 2005; Rohrlich 2007; Muller
2007), it must be noted that the Maxwell--Lorentz equations (\ref{eq:MLE1})--(\ref{eq:MLE5}),
exactly in this form, have \emph{no} solution. The reason is the following.
In the Lorentz equation of motion (\ref{eq:E1'}), a small but extended
particle can be described with a good approximation by one single
specific passive electric charge $\pi^{b}$ and one single trajectory
$\mathbf{r}^{b}\left(t\right)$. In contrast, however, a similar {}``idealization''
in the source densities (\ref{eq:(E3)1})--(\ref{eq:(E3)2}) leads
to singularities; the field is singular at precisely the points where
the coupling happens: on the trajectory of the particle. 

The generally accepted answer to this problem is that (\ref{eq:(E3)1})--(\ref{eq:(E3)2})
should not be taken literally. Due to the inner structure of the particle,
the real source densities are some {}``smoothed out'' Dirac deltas.
Instead of (\ref{eq:(E3)1})--(\ref{eq:(E3)2}), therefore, we have
some more general equations
\begin{eqnarray}
\left[\varrho(\mathbf{r},t)\right] & = & \mathcal{R}^{b}\left[\mathbf{r}^{b}(t)\right]\label{eq:struktura1}\\
\left[\mathbf{j}(\mathbf{r},t)\right] & = & \mathcal{J}^{b}\left[\mathbf{r}^{b}(t)\right]\label{eq:struktura2}
\end{eqnarray}
where $\mathcal{R}^{b}$ and $\mathcal{J}^{b}$ are, generally non-linear,
operators providing functional relationships between the particle's
trajectory $\left[\mathbf{r}^{b}(t)\right]$ and the source density
functions $\left[\varrho(\mathbf{r},t)\right]$ and $\left[\mathbf{j}(\mathbf{r},t)\right]$.
(Notice that (\ref{eq:(E3)1})--(\ref{eq:(E3)2}) serve as example
of such equations.) The concrete forms of equations (\ref{eq:struktura1})--(\ref{eq:struktura2})
are determined by the physical laws of the internal world of the particle---which
are, supposedly, outside of the scope of ED. At this level of generality,
the only thing we can say is that, for a {}``point-like'' (localized)
particle, equations (\ref{eq:struktura1})--(\ref{eq:struktura2})
must be something very close to---but not identical with---equations
(\ref{eq:(E3)1})--(\ref{eq:(E3)2}). With this explanation, for the
sake of simplicity we leave the Dirac deltas in the equations. Also,
in some of our statements and calculations the Dirac deltas are essentially
used; for example, (E3) and, partly, Theorem~\ref{thm:-E2'} and
\ref{thm:superpos'} would not be true without the exact point-like
source densities (\ref{eq:(E3)1})--(\ref{eq:(E3)2}). But a little
reflection shows that the statements in question remain approximately
true if the particles are approximately point-like, that is, if equations
(\ref{eq:struktura1})--(\ref{eq:struktura2}) are close enough to
equations (\ref{eq:(E3)1})--(\ref{eq:(E3)2}). To be noted that what
is actually essential in (\ref{eq:(E3)1})--(\ref{eq:(E3)2}) is not
the point-likeness of the particle, but its stability; no matter how
the system moves, it remains a localized object. \hfill{} $\lrcorner$

\section{Operational definitions of electrodynamical quantities in $K'$}

So far we have only considered ED in a single frame of reference $K$.
Now we turn to the question of how a moving observer describes the
same phenomena in $K'$. The observed phenomena are the same, but
the measuring equipments by means of which the phenomena are observed
are not entirely the same; instead of being at rest in $K$, they
are co-moving with\emph{ $K'$}.

Accordingly, we will repeat the operational definitions (D0)--(D4)
with the following differences:
\begin{enumerate}
\item The {}``rest test particles'' will be at rest relative to reference
frame $K'$, that is, \emph{in motion with velocity $\mathbf{V}$}
relative to $K$.
\item The measuring equipments by means of which the kinematical quantities
are ascertained---say, the measuring rods and clocks---will be at
rest relative to $K'$, that is, \emph{in motion with velocity $\mathbf{V}$}
relative to $K$. In other words, kinematical quantities $t,\mathbf{r},\mathbf{v},\mathbf{a}$
in definitions (D0)--(D4) will be \emph{replaced with}---not expressed
in terms of--- $t',\mathbf{r}',\mathbf{v}',\mathbf{a}'$.
\end{enumerate}

\paragraph*{Definition~(D0')}

Particle $e$ is called \emph{(test particle)'} if for all $\mathbf{r}'$
and $t'$
\begin{equation}
\mathbf{v}'^{e}\left(t'\right)\biggl|_{\mathbf{r}'^{e}\left(t'\right)=\mathbf{r}'}=\mathbf{v}'^{etalon}\left(t'\right)\biggl|_{\mathbf{r}'^{etalon}\left(t'\right)=\mathbf{r}'}
\end{equation}
implies
\begin{equation}
\mathbf{a}'^{e}\left(t'\right)\biggl|_{\mathbf{r}'^{e}\left(t'\right)=\mathbf{r}'}=\mathbf{a}'^{etalon}\left(t'\right)\biggl|_{\mathbf{r}'^{etalon}\left(t'\right)=\mathbf{r}'}
\end{equation}
\medskip{}

\noindent A (test particle)' $e$ moving with velocity $\mathbf{V}$
relative to $K$ is at rest relative to $K'$, that is, $\mathbf{v}'^{e}=0$.
Accordingly:

\paragraph*{Definition~(D1')}

\noindent \emph{(Electric field strength)'} at point $\mathbf{r}'$
and time $t'$ is defined as the acceleration of an arbitrary (test
particle)' $e$, such that $\mathbf{r}'^{e}(t)=\mathbf{r}'$ and $\mathbf{v}'^{e}(t')=0$:
\begin{equation}
\mathbf{E}'\left(\mathbf{r}',t'\right)\overset{^{def}}{=}\left.\mathbf{a}'^{e}(t')\right|_{\mathbf{r}'^{e}(t')=\mathbf{r}';\,\mathbf{v}'^{e}(t')=0}\label{eq:E-def-vesszo}
\end{equation}
\medskip{}

\noindent Similarly, (magnetic field strength)' is defined by means
of how the acceleration $\mathbf{a}'^{e}$ of a rest (test particle)'---rest,
of course, relative to $K'$---changes with a small perturbation of
its state of motion, that is, if an infinitesimally small velocity
$\mathbf{v}'^{e}$ is imparted to the particle. Just as in (D2), let
$\delta'\subset\mathbb{R}^{3}$ be an arbitrary infinitesimal neighborhood
of $0\in\mathbb{R}^{3}$. We define the following function:
\begin{eqnarray}
\mathbf{U}'^{\mathbf{r}',t'} & : & \mathbb{R}^{3}\supset\delta'\rightarrow\mathbb{R}^{3}\nonumber \\
 &  & \mathbf{U}'^{\mathbf{r}',t'}(\mathbf{v}')\overset{^{def}}{=}\left.\mathbf{a}'^{e}(t')\right|_{\mathbf{r}'^{e}(t')=\mathbf{r}';\,\mathbf{v}'^{e}(t')=\mathbf{v}'}\label{eq:U-def-vesszo}
\end{eqnarray}

\paragraph*{Definition~(D2')}

\noindent \emph{(Magnetic field strength)'} at point $\mathbf{r}'$
and time $t'$ is

\begin{equation}
\mathbf{B}'(\mathbf{r}',t')\overset{^{def}}{=}\left.\left(\begin{array}{l}
\partial_{v'_{z}}U'{}_{y}^{\mathbf{r}',t'}\\
\partial_{v'_{x}}U'{}_{z}^{\mathbf{r}',t'}\\
\partial_{v'_{y}}U'{}_{x}^{\mathbf{r}',t'}
\end{array}\right)\right|_{\mathbf{v}'=0}\label{eq:B-def-vesszo}
\end{equation}

\paragraph*{Definition~(D3')}

\begin{eqnarray}
\varrho'\left(\mathbf{r}',t'\right) & \overset{^{def}}{=} & \nabla\cdot\mathbf{E}'\left(\mathbf{r}',t'\right)\label{eq:ME1'}\\
\mathbf{j}'\left(\mathbf{r}',t'\right) & \overset{^{def}}{=} & c^{2}\nabla\times\mathbf{B}'\left(\mathbf{r}',t'\right)-\partial_{t'}\mathbf{E}'\left(\mathbf{r}',t'\right)\label{eq:ME2'}
\end{eqnarray}
are called \emph{(active electric charge density)'} and \emph{(active
electric current density)',} respectively. \medskip{}

\noindent Of course, we have:
\begin{thm}
~
\begin{equation}
\partial_{t'}\varrho'\left(\mathbf{r}',t'\right)+\nabla\cdot\mathbf{j}'\left(\mathbf{r}',t'\right)=0\label{eq:kontinuitas'}
\end{equation}

\end{thm}

\paragraph*{Definition (D4')}

A particle is called \emph{(charged point-particle)'} of \emph{(specific
passive electric charge)'} $\pi'^{b}$ and of \emph{(active electric
charge)'} $\alpha'^{b}$ if the following is true:
\begin{enumerate}
\item It satisfies the relativistic Lorentz equation,

\noindent 
\begin{eqnarray}
\gamma\left(\mathbf{v}'^{b}\left(t'\right)\right)\mathbf{a}'^{b}(t') & = & \pi'^{b}\left\{ \mathbf{E}'\left(\mathbf{r}'^{b}\left(t'\right),t'\right)+\mathbf{v}'^{b}\left(t'\right)\times\mathbf{B}'\left(\mathbf{r}'^{b}\left(t'\right),t'\right)\right.\nonumber \\
 &  & \left.-c^{-2}\mathbf{v}'^{b}\left(t'\right)\left(\mathbf{v}'^{b}\left(t'\right)\mathbf{\cdot E}'\left(\mathbf{r}'^{b}\left(t'\right),t'\right)\right)\right\} \label{eq:E1'-vesszo}
\end{eqnarray}

\item \noindent If it is the only particle whose worldline intersects a
given space-time region $\Lambda'$, then for all $(\mathbf{r}',t')\in\Lambda'$
the (source densities)' are of the following form: 
\begin{eqnarray}
\varrho'\left(\mathbf{r}',t'\right) & = & \alpha'^{b}\delta\left(\mathbf{r}'-\mathbf{r}'^{b}\left(t'\right)\right)\label{eq:(E3)1-vesszo}\\
\mathbf{j}'\left(\mathbf{r}',t'\right) & = & \alpha'^{b}\delta\left(\mathbf{r}'-\mathbf{r}'^{b}\left(t'\right)\right)\mathbf{v}'^{b}\left(t'\right)\label{eq:(E3)2-vesszo}
\end{eqnarray}

\end{enumerate}
where $\mathbf{r}'^{b}\left(t'\right)$, $\mathbf{v}'^{b}\left(t'\right)$
and $\mathbf{a}'^{b}\left(t'\right)$ is the particle's position,
velocity and acceleration in $K'$. The ratio $\mu'^{b}\overset{^{def}}{=}\alpha'^{b}/\pi'^{b}$
is called the \emph{(electric inertial rest mass)'} of the particle.\medskip{}

\R\label{meg:5}It is worthwhile to make a few remarks about some
epistemological issues:
\begin{lyxlist}{00.00.0000}
\item [{(a)}] The physical quantities defined in (D1)--(D4) \emph{differ}
from the physical quantities defined in (D1')--(D4'), simply because
the physical situation in which a test particle is at rest relative
to $K$ differs from the one in which it is co-moving with $K'$ with
velocity $\mathbf{V}$ relative to $K$; and, as we know \emph{from
the laws of ED in $K$}, this difference really matters.

Someone might object that if this is so then any two instances of
the same measurement must be regarded as measurements of different
physical quantities. For, if the difference in the test particle's
velocity is enough reason to say that the two operations determine
two different quantities, then, by the same token, two operations
must be regarded as different operations---and the corresponding quantities
as different physical quantities---if the test particle is at different
points of space, or the operations simply happen at different moments
of time. And this consequence, the objection goes, seems to be absurd:
if it were true, then science would not be possible, because we would
not have the power to make law-like assertions at all; therefore we
must admit that empiricism fails to explain how natural laws are possible,
and, as many argue, science cannot do without metaphysical pre-assumptions.

Our response to such an objections is the following. First, concerning
the general epistemological issue, we believe, nothing disastrous
follows from admitting that two phenomena observed at different place
or at different time \emph{are} distinct. And if they are stated as
instances of the same phenomenon, this statement is not a logical
or metaphysical necessity---derived from some logical/metaphysical
pre-assumptions---but an ordinary scientific hypothesis obtained by
induction and confirmed or disconfirmed together with the \emph{whole}
scientific theory. In fact, this is precisely the case with respect
to the definitions of the fundamental electrodynamical quantities.
For example, definition (D1) is in fact a family of definitions each
belonging to a particular situation individuated by the space-time
locus $(\mathbf{r},t)$.\\
Second, in this paper, we must emphasize again, the question of operational
definitions of electrodynamical quantities first of all emerges not
from an epistemological context, but from the context of the \emph{inner
consistency} of our theories, in answering questions (Q3) and (Q4).
In the next section, all the results of the measurement operations
defined in (D1')--(D4') will be predicted from the laws of ED in $K$.
And, ED itself says that some differences in the conditions are relevant
from the point of view of the measured accelerations of the test particles,
some others are not; some of the originally distinct quantities are
contingently equal, some others not.

\item [{(b)}] From a mathematical point of view, both (D0)--(D4) and (D0')--(D4')
are definitions. However, while the choice of the \emph{etalon} test
particle and definitions (D0)--(D4) are entirely \emph{conventional},
there is no additional conventionality in (D0')--(D4'). The way in
which we define the electrodynamical quantities in inertial frame
$K'$ automatically follows from (D0)--(D4) and from the question
we would like to answer, namely, whether the RP holds for ED; since
the principle is about {}``quantities obtained by the same operational
procedures with the same measuring equipments when they are co-moving
with\emph{ $K'$}''.\emph{ }
\item [{(c)}] In fact, one of the constituents of the concepts defined
in $K'$ is not determined by the operational definitions in $K$.
Namely, the notion of {}``the same operational procedures with the
same measuring equipments when they are co-moving with\emph{ $K'$}'',
that is, the notion of $M_{\mathbf{V}}$ applied for the measuring
operation and the measuring equipments. This is however not an additional
freedom of conventionality, but a simple vagueness in our physical
theories in $K$. In any event, in our case, the notion of the only
moving measuring device, that is, the notion of {}``a test particle
at rest relative to $K'$'' is quite clear.\hfill{} $\lrcorner$
\end{lyxlist}

\section{Observations of moving observer\label{sec:Observations-of-moving}}

Now we have another collection of operationally defined notions, $\mathbf{E}',\mathbf{B}',$$\varrho',\mathbf{j}'$,
the concept of (charged point-particle)' defined in the primed terms,
and its properties $\pi',\alpha'$ and $\mu'$. Normally, one should
investigate these quantities experimentally and collect new empirical
facts about both the relationships between the primed quantities and
about the relationships between the primed quantities and the ones
defined in (D1)--(D4). In contrast, we will continue our analysis
in another way; following the {}``Lorentzian pedagogy'', we will
determine from the laws of physics in $K$ what an observer co-moving
with $K'$ should observe. In fact, with this method, we will answer
our question (Q4), whether the textbook transformation rules, derived
from the RP, are compatible with the laws of ED in a single frame
of reference. We will also see whether the basic equations \eqref{eq:MLE1}--\eqref{eq:MLE5}
are covariant against these transformations.

Throughout the theorems below, it is important that when we compare,
for example, $\mathbf{E}\left(\mathbf{r},t\right)$ with $\mathbf{E}'(\mathbf{r}',t')$,
we compare the values of the fields \emph{in one and the same event},
that is, we compare $\mathbf{E}\left(\mathbf{r}(A),t(A)\right)$ with
$\mathbf{E}'\left(\mathbf{r}'(A),t'(A)\right)$. For the sake of brevity,
however, we omit the indication of this fact.

The first theorem trivially follows from the fact that the Lorentz
transformations of the kinematical quantities are one-to-one: 
\begin{thm}
\label{thm:test=00003Dtest'}A particle is a (test particle)' if and
only if it is a test particle.
\end{thm}
\noindent Consequently, we have many enough (test particles)' for
definitions (D1')--(D4'); and each is a charged point-particle satisfying
the Lorentz equation \eqref{eq:E1'} with specific passive electric
charge $\pi=1$. 
\begin{thm}
~\label{thm:E'E}
\begin{eqnarray}
E'_{x} & = & E_{x}\label{eq:E'x}\\
E'_{y} & = & \gamma\left(E_{y}-VB_{z}\right)\label{eq:E'y}\\
E'_{z} & = & \gamma\left(E_{z}+VB_{y}\right)\label{eq:E'z}
\end{eqnarray}
\end{thm}
\begin{proof}
\noindent When the (test particle)' is at rest relative to $K'$,
it is moving with velocity $\mathbf{v}^{e}=\left(V,0,0\right)$ relative
to $K$. From \eqref{eq:E1'} (with $\pi=1$) we have
\begin{eqnarray}
a_{x}^{e} & = & \gamma^{-3}E_{x}\label{eq:TTgyors1}\\
a_{y}^{e} & = & \gamma^{-1}\left(E_{y}-VB_{z}\right)\\
a_{z}^{e} & = & \gamma^{-1}\left(E_{z}+VB_{y}\right)\label{eq:TTgyors3}
\end{eqnarray}
Applying \eqref{eq:gyorsulas1}--\eqref{eq:gyorsulas3}, we can calculate
the acceleration $\mathbf{a}'^{e}$ in $K'$, and, accordingly, we
find
\begin{eqnarray}
E'_{x} & = & a{}_{x}^{\prime e}=\gamma^{3}a_{x}^{e}=E_{x}\\
E'_{y} & = & a{}_{y}^{\prime e}=\gamma^{2}a_{y}^{e}=\gamma\left(E_{y}-VB_{z}\right)\\
E'_{z} & = & a{}_{z}^{\prime e}=\gamma^{2}a_{z}^{e}=\gamma\left(E_{z}+VB_{y}\right)
\end{eqnarray}
 \end{proof}
\begin{thm}
\label{thm:B'B}
\begin{eqnarray}
B'_{x} & = & B{}_{x}\label{eq:Bx'}\\
B'_{y} & = & \gamma\left(B_{y}+c^{-2}VE_{z}\right)\\
B'_{z} & = & \gamma\left(B_{z}-c^{-2}VE_{y}\right)\label{eq:Bz'}
\end{eqnarray}
\end{thm}
\begin{proof}
\noindent Consider for instance $B'_{x}$. By definition, 
\begin{equation}
B'_{x}=\left.\partial_{v'_{z}}U'{}_{y}^{\mathbf{r}',t'}\right|_{\mathbf{v}'=0}\label{eq:Bxdef}
\end{equation}
According to \eqref{eq:U-def-vesszo}, the value of $U'{}_{y}^{\mathbf{r}',t'}(\mathbf{v}')$
is equal to 
\begin{equation}
\left.a'{}_{y}^{e}\right|_{\mathbf{r}'^{e}(t')=\mathbf{r}';\,\mathbf{v}'^{e}(t')=\mathbf{v}'}
\end{equation}
that is, the $y$-component of the acceleration of a (test particle)'
$e$ in a situation in which $\mathbf{r}'^{e}(t')=\mathbf{r}'$ and
$\mathbf{v}'^{e}(t')=\mathbf{v}'$. Accordingly, in order to determine
the partial derivative \eqref{eq:Bxdef} we have to determine 
\begin{equation}
\left.\frac{d}{dw}\right|_{w=0}\left(\left.a'{}_{y}^{e}\right|_{\mathbf{r}'^{e}(t')=\mathbf{r}';\,\mathbf{v}'^{e}(t')=\left(0,0,w\right)}\right)
\end{equation}
Now, according to \eqref{eq:sebesseg1}, condition $\mathbf{v}'^{e}=\left(0,0,w\right)$
corresponds to 
\begin{equation}
\mathbf{v}^{e}=\left(V,0,\gamma^{-1}w\right)
\end{equation}
Substituting this velocity into \eqref{eq:E1'}, we have:
\begin{equation}
a_{y}^{e}=\sqrt{1-\frac{V^{2}+w{}^{2}\gamma^{-2}}{c^{2}}}\left(E_{y}+w\gamma^{-1}B_{x}-VB_{z}\right)\label{eq:hivatkozott-a}
\end{equation}
Applying \eqref{eq:gyorsulas4}, one finds:
\begin{eqnarray}
a{}_{y}^{\prime e} & = & \gamma^{2}a_{y}^{e}=\gamma^{2}\sqrt{1-\frac{V^{2}+w{}^{2}\gamma^{-2}}{c^{2}}}\left(E_{y}+w\gamma^{-1}B_{x}-VB_{z}\right)\nonumber \\
 & = & \frac{\gamma}{\gamma(w)}\left(E_{y}+w\gamma^{-1}B_{x}-VB_{z}\right)\label{eq:gyorsulasB}
\end{eqnarray}
Differentiating with respect to $w$ at $w=0$, we obtain
\begin{equation}
B'_{x}=B{}_{x}
\end{equation}
The other components can be obtained in the same way.\end{proof}
\begin{thm}
\label{thm:aramtrafo}~
\begin{eqnarray}
\varrho' & = & \gamma\left(\varrho-c^{-2}Vj_{x}\right)\label{eq:surusegtarfo1}\\
j'_{x} & = & \gamma\left(j_{x}-V\varrho\right)\\
j'_{y} & = & j_{y}\\
j'_{z} & = & j_{z}\label{eq:surusegtrafo4}
\end{eqnarray}
\end{thm}
\begin{proof}
Substituting $\mathbf{E}'$ and $\mathbf{B}'$ with \eqref{eq:E'x}--\eqref{eq:E'z}
and \eqref{eq:Bx'}--\eqref{eq:Bz'}, $\mathbf{r}$ and $t$ with
the inverse of \eqref{eq:LT1}--\eqref{eq:LT4}, then differentiating
the composite function and taking into account \eqref{eq:ME1}--\eqref{eq:ME2},
we get \eqref{eq:surusegtarfo1}--\eqref{eq:surusegtrafo4}.\end{proof}
\begin{thm}
\label{thm:-E2'}A particle $b$ is charged point-particle of specific
passive electric charge\emph{ }$\pi{}^{b}$ and of active electric\emph{
}charge $\alpha{}^{b}$ if and only if it is a (charged point-particle)'
of (specific passive electric charge)'\emph{ }$\pi'^{b}$ and of (active
electric\emph{ }charge)' $\alpha'{}^{b}$, such that $\pi'^{b}=\pi{}^{b}$
and $\alpha'{}^{b}=\alpha^{b}$. \end{thm}
\begin{proof}
First we prove \eqref{eq:E1'-vesszo}. For the sake of simplicity,
we will verify this in case of $\mathbf{v}{}^{\prime b}=\left(0,0,w\right)$.
We can use \eqref{eq:hivatkozott-a}: 
\begin{eqnarray}
a_{y}^{b} & = & \pi^{b}\sqrt{1-\frac{V^{2}+w{}^{2}\gamma^{-2}}{c^{2}}}\left(E_{y}+w\gamma^{-1}B_{x}-VB_{z}\right)
\end{eqnarray}
From \eqref{eq:gyorsulas4}, \eqref{eq:E'y}, \eqref{eq:Bx'}, and
\eqref{eq:Bz'} we have 
\begin{eqnarray}
a{}_{y}^{\prime b} & = & \pi^{b}\gamma(w)^{-1}\left(E'_{y}+wB'_{x}\right)\nonumber \\
 & = & \left[\pi^{b}\gamma\left(\mathbf{v}'^{b}\right)^{-1}\left(\mathbf{E}'-c^{-2}v'^{b}\left(\mathbf{v}'^{b}\mathbf{\cdot E}'\right)+\mathbf{v}'^{b}\times\mathbf{B}'\right)\right]_{y}\Biggl|_{\mathbf{v'}^{b}=\left(0,0,w\right)}
\end{eqnarray}
Similarly, 
\begin{eqnarray}
a{}_{x}^{\prime b} & = & \pi^{b}\gamma(w)^{-1}\left(E'_{x}-wB'_{y}\right)\nonumber \\
 & = & \left[\pi^{b}\gamma\left(\mathbf{v}'^{b}\right)^{-1}\left(\mathbf{E}'-c^{-2}\mathbf{v}'^{b}\left(\mathbf{v}'^{b}\mathbf{\cdot E}'\right)+\mathbf{v}'^{b}\times\mathbf{B}'\right)\right]_{x}\Biggl|_{\mathbf{v'}^{b}=\left(0,0,w\right)}\\
a{}_{z}^{\prime b} & = & \pi^{b}\gamma(w)^{-3}E'_{z}\nonumber \\
 & = & \left[\pi^{b}\gamma\left(\mathbf{v}'^{b}\right)^{-1}\left(\mathbf{E}'-c^{-2}\mathbf{v}'^{b}\left(\mathbf{v}'^{b}\mathbf{\cdot E}'\right)+\mathbf{v}'^{b}\times\mathbf{B}'\right)\right]_{z}\Biggl|_{\mathbf{v'}^{b}=\left(0,0,w\right)}
\end{eqnarray}
That is, \eqref{eq:E1'-vesszo} is satisfied, indeed.

In the second part, we show that \eqref{eq:(E3)1-vesszo}--\eqref{eq:(E3)2-vesszo}
are nothing but \eqref{eq:(E3)1}--\eqref{eq:(E3)2} expressed in
terms of $\mathbf{r}',t',\varrho'$ and $\mathbf{j}'$, with $\alpha^{'b}=\alpha^{b}$. 

It will be demonstrated for a particle of trajectory $\mathbf{r}'^{b}\left(t'\right)=\left(wt',0,0\right)$.
Applying \eqref{eq:sebesseg3}, \eqref{eq:(E3)1}--\eqref{eq:(E3)2}
have the following forms:
\begin{eqnarray}
\varrho{}\left(\mathbf{r},t\right) & = & \alpha^{b}\delta\left(x-\beta t\right)\delta\left(y\right)\delta\left(z\right)\\
\mathbf{j}{}\left(\mathbf{r},t\right) & = & \alpha^{b}\delta\left(x-\beta t\right)\delta\left(y\right)\delta\left(z\right)\left(\begin{array}{c}
\beta\\
0\\
0
\end{array}\right)
\end{eqnarray}
where $\beta=\frac{w+V}{1+c^{-2}wV}$. $\mathbf{r},t,\varrho$ and
$\mathbf{j}$ can be expressed with the primed quantities by applying
the inverse of \eqref{eq:LT1}--\eqref{eq:LT4} and \eqref{eq:surusegtarfo1}--\eqref{eq:surusegtrafo4}:
\begin{eqnarray}
\gamma\left(\varrho'{}\left(\mathbf{r}',t'\right)+c^{-2}Vj'{}_{x}\left(\mathbf{r}',t'\right)\right) & = & \alpha^{b}\delta\left(\gamma\left(x'+Vt'-\beta\left(t'+c^{-2}Vx'\right)\right)\right)\nonumber \\
 &  & \times\,\delta\left(y'\right)\delta\left(z'\right)\\
\gamma\left(j'{}_{x}\left(\mathbf{r}',t'\right)+V\varrho'{}\left(\mathbf{r}',t'\right)\right) & = & \alpha^{b}\delta\left(\gamma\left(x'+Vt'-\beta\left(t'+c^{-2}Vx'\right)\right)\right)\nonumber \\
 &  & \times\,\delta\left(y'\right)\delta\left(z'\right)\beta\\
j'{}_{y}\left(\mathbf{r}',t'\right) & = & 0\\
j'{}_{z}\left(\mathbf{r}',t'\right) & = & 0
\end{eqnarray}
One can solve this system of equations for $\varrho'$ and $j'_{x}$:
\begin{eqnarray}
\varrho'\left(\mathbf{r}',t'\right) & = & \alpha^{b}\delta\left(x'-wt'\right)\delta\left(y'\right)\delta\left(z'\right)\label{eq:suruseg-trafo_reszecske1}\\
\mathbf{j}'\left(\mathbf{r}',t'\right) & = & \alpha^{b}\delta\left(x'-wt'\right)\delta\left(y'\right)\delta\left(z'\right)\left(\begin{array}{c}
w\\
0\\
0
\end{array}\right)\label{eq:suruseg-trafo_reszecske2}
\end{eqnarray}
\end{proof}
\begin{thm}
~
\begin{eqnarray}
\nabla\cdot\mathbf{B}'\left(\mathbf{r}',t'\right) & = & 0\label{eq:ME3'}\\
\nabla\times\mathbf{E}'\left(\mathbf{r}',t'\right)+\partial_{t'}\mathbf{B}'\left(\mathbf{r}',t'\right) & = & 0\label{eq:ME4'}
\end{eqnarray}
\end{thm}
\begin{proof}
Expressing \eqref{eq:ME3}--\eqref{eq:ME4} in terms of $\mathbf{r}',t',\mathbf{E}'$
and $\mathbf{B}'$ by means of \eqref{eq:LT1}--\eqref{eq:LT4}, \eqref{eq:E'x}--\eqref{eq:E'z}
and \eqref{eq:Bx'}--\eqref{eq:Bz'}, we have
\begin{eqnarray}
\nabla\cdot\mathbf{B}'-c^{-2}V\left(\nabla\times\mathbf{E}'+\partial_{t'}\mathbf{B}'\right)_{x} & = & 0\\
\left(\nabla\times\mathbf{E}'+\partial_{t'}\mathbf{B}'\right)_{x}-V\nabla\cdot\mathbf{B}' & = & 0\\
\left(\nabla\times\mathbf{E}'+\partial_{t'}\mathbf{B}'\right)_{y} & = & 0\\
\left(\nabla\times\mathbf{E}'+\partial_{t'}\mathbf{B}'\right)_{z} & = & 0
\end{eqnarray}
which is equivalent to \eqref{eq:ME3'}--\eqref{eq:ME4'}, indeed.\end{proof}
\begin{thm}
\label{thm:superpos'}If $b_{1}$, $b_{2}$,..., $b_{n}$ are the
only particles whose worldlines intersect a given space-time region
$\Lambda'$, then for all $\left(\mathbf{r}',t'\right)\in\Lambda'$
the (source densities)' are:
\begin{eqnarray}
\varrho'\left(\mathbf{r}',t'\right) & = & \sum_{i=1}^{n}\alpha^{b_{i}}\delta\left(\mathbf{r}'-\mathbf{r}'^{b_{i}}\left(t'\right)\right)\label{eq:add1'}\\
\mathbf{j}'\left(\mathbf{r}',t'\right) & = & \sum_{i=1}^{n}\alpha^{b_{i}}\delta\left(\mathbf{r}'-\mathbf{r}'^{b_{i}}\left(t'\right)\right)\mathbf{v}'^{b_{i}}\left(t'\right)\label{eq:add2'}
\end{eqnarray}
\end{thm}
\begin{proof}
Due to Theorem~\ref{thm:-E2'}, each (charged point-particle)' is
a charged point-particle with $\alpha^{'b}=\alpha^{b}$. Therefore,
we only need to prove that equations \eqref{eq:add1'}--\eqref{eq:add2'}
amount to \eqref{eq:add1}--\eqref{eq:add2} expressed in the primed
variables. On the left hand side of \eqref{eq:add1}--\eqref{eq:add2},
$\varrho$ and $\mathbf{j}$ can be expressed by means of the inverse
of \eqref{eq:surusegtarfo1}--\eqref{eq:surusegtrafo4}; on the right
hand side, we take $\alpha^{'b}=\alpha^{b}$, and apply the inverse
of \eqref{eq:LT1}--\eqref{eq:LT4}, just as in the derivation of
\eqref{eq:suruseg-trafo_reszecske1}--\eqref{eq:suruseg-trafo_reszecske2}.
From the above, we obtain: 

\begin{eqnarray}
\varrho'\left(\mathbf{r}',t'\right)+c^{-2}Vj'{}_{x}\left(\mathbf{r}',t'\right) & = & \sum_{i=1}^{n}\alpha^{b_{i}}\delta\left(\mathbf{r}'-\mathbf{r}'^{b_{i}}\left(t'\right)\right)\nonumber \\
 &  & +c^{-2}V\sum_{i=1}^{n}\alpha^{b_{i}}\delta\left(\mathbf{r}'-\mathbf{r}'^{b_{i}}\left(t'\right)\right)v_{x}'^{b_{i}}\left(t'\right)\\
j'{}_{x}\left(\mathbf{r}',t'\right)+V\varrho'\left(\mathbf{r}',t'\right) & = & \sum_{i=1}^{n}\alpha^{b_{i}}\delta\left(\mathbf{r}'-\mathbf{r}'^{b_{i}}\left(t'\right)\right)v_{x}'^{b_{i}}\left(t'\right)\nonumber \\
 &  & +V\sum_{i=1}^{n}\alpha^{b_{i}}\delta\left(\mathbf{r}'-\mathbf{r}'^{b_{i}}\left(t'\right)\right)\\
j'{}_{y}\left(\mathbf{r}',t'\right) & = & \sum_{i=1}^{n}\alpha^{b_{i}}\delta\left(\mathbf{r}'-\mathbf{r}'^{b_{i}}\left(t'\right)\right)v_{y}'^{b_{i}}\left(t'\right)\\
j'{}_{z}\left(\mathbf{r}',t'\right) & = & \sum_{i=1}^{n}\alpha^{b_{i}}\delta\left(\mathbf{r}'-\mathbf{r}'^{b_{i}}\left(t'\right)\right)v_{z}'^{b_{i}}\left(t'\right)
\end{eqnarray}
Solving these linear equations for $\varrho'$ and $\mathbf{j}'$
we obtain \eqref{eq:add1'}--\eqref{eq:add2'}.
\end{proof}
Combining the results we obtained in Theorems~\ref{thm:-E2'}--\ref{thm:superpos'},
we have
\begin{eqnarray}
\nabla\cdot\mathbf{E}'\left(\mathbf{r}',t'\right) & = & \sum_{i=1}^{n}\alpha^{b_{i}}\delta\left(\mathbf{r}'-\mathbf{r}'^{b_{i}}\left(t'\right)\right)\label{eq:MLE1'}\\
c^{2}\nabla\times\mathbf{B}'\left(\mathbf{r}',t'\right)-\partial_{t'}\mathbf{E}'\left(\mathbf{r}',t'\right) & = & \sum_{i=1}^{n}\alpha^{b_{i}}\delta\left(\mathbf{r}'-\mathbf{r}'^{b_{i}}\left(t'\right)\right)\mathbf{v}'^{b_{i}}\left(t'\right)\label{eq:MLE2'}\\
\nabla\cdot\mathbf{B}'\left(\mathbf{r}',t'\right) & = & 0\label{eq:MLE3'}\\
\nabla\times\mathbf{E}'\left(\mathbf{r}',t'\right)+\partial_{t'}\mathbf{B}'\left(\mathbf{r}',t'\right) & = & 0\label{eq:MLE4'}\\
\gamma\left(\mathbf{v}'^{b_{i}}\left(t'\right)\right)\mathbf{a}'^{b_{i}}(t') & = & \pi'^{b_{i}}\Biggl\{\mathbf{E}'\left(\mathbf{r}'^{b_{i}}\left(t'\right),t'\right)\nonumber \\
 &  & +\mathbf{v}'^{b_{i}}\left(t'\right)\times\mathbf{B}'\left(\mathbf{r}'^{b_{i}}\left(t'\right),t'\right)\nonumber \\
 &  & -\mathbf{v}'^{b_{i}}\left(t'\right)\frac{\mathbf{v}'^{b_{i}}\left(t'\right)\mathbf{\cdot E}'\left(\mathbf{r}'^{b_{i}}\left(t'\right),t'\right)}{c^{2}}\Biggr\}\,\,\,\,\,\,\,\,\label{eq:MLE5'}\\
 &  & \,\,\,\,\,\,\,\,\,\,\,\,\,\,\,\,\,\,\,\,\,\,\,\,\,\,\,\,\,\,\,\,\,\,\,\,\,(i=1,2,\ldots n)\nonumber 
\end{eqnarray}

\section{Are the textbook transformation rules consistent with the laws of
ED in a single frame of reference?\label{sec:Are-the-textbook} }

Now, everything is at hand to declare that the textbook transformation
rules for electrodynamical quantities, routinely derived from the
\emph{presumed} covariance of the Maxwell equations, are in fact true,
at least in the sense that they are derivable from the laws of ED
in a single frame of reference, including---it must be emphasized---the
precise operational definitions of the quantities in question. For,
Theorems~\ref{thm:E'E} and \ref{thm:B'B} show the well-known transformation
rules for the field variables. What Theorem~\ref{thm:aramtrafo}
asserts is nothing but the well-known transformation rule for charge
density and current density. Finally, Theorem~\ref{thm:-E2'} shows
that a particle's electric specific passive charge, active charge
and electric rest mass are invariant scalars.

At this point, having ascertained the transformation rules, we can
declare that equations (\ref{eq:MLE1'})--(\ref{eq:MLE5'}) are nothing
but $T_{\mathbf{V}}(\mathcal{E})$ (in coordinates, of course), where
$\mathcal{E}$ stands for the equations (\ref{eq:MLE1})--(\ref{eq:MLE5}).
At the same time, (\ref{eq:MLE1'})--(\ref{eq:MLE5'}) are manifestly
equal to $P_{\mathbf{V}}(\mathcal{E})$. Therefore, we proved that
the Maxwell--Lorentz equations are covariant against the transformations
of the kinematical and electrodynamical quantities. In fact, we proved
more: 
\begin{itemize}
\item The Lorentz equation of motion \eqref{eq:MLE5} is covariant separately.
\item The four Maxwell equations (\ref{eq:MLE1})--(\ref{eq:MLE4}) constitute
a covariant set of equations, separately from \eqref{eq:MLE5}.
\item (\ref{eq:MLE1})--(\ref{eq:MLE2}) constitute a covariant set of equations,
separately.
\item (\ref{eq:MLE3})--(\ref{eq:MLE4}) constitute a covariant set of equations,
separately.
\end{itemize}
As we pointed out in Remark~\ref{meg:kovariancia-uj}, none of these
statements follows automatically from the fact that (\ref{eq:MLE1})--(\ref{eq:MLE5})
is a covariant system of equations.

\R\label{meg:6}The fact that the proper calculation of the transformation
rules for the field strengths and for the source densities leads to
the familiar textbook transformation rules hinges on the \emph{relativistic}
version of the Lorentz equation, in particular, on the {}``relativistic
mass-formula''. Without factor $\gamma\left(\mathbf{v}{}^{b}\right)$
in (\ref{eq:MLE5}), the proper transformation rules were different
and the Maxwell equations were not covariant---against the proper
transformations. \hfill{} $\lrcorner$

\R\label{meg:7}This is not the place to review the various versions
of the textbook derivation of the transformation rules for electrodynamical
quantities, nevertheless, a few remarks seem necessary. Among those
with which we are acquainted, there are basically two major branches,
and both are problematic. The first version follows Einstein's 1905
paper: 
\begin{lyxlist}{00.00.0000}
\item [{(1a)}] The transformation rules of electric and magnetic field
strengths are derived from the presumption of the covariance of the
homogeneous (with no sources) Maxwell equations.
\item [{(1b)}] The transformation rules of source densities are derived
from the transformations of the field variables. 
\item [{(1c)}] From the transformation rules of charge and current densities,
it is derived that electric charge is an invariant scalar. 
\end{lyxlist}
The second version is this: 
\begin{lyxlist}{00.00.0000}
\item [{(2a)}] The transformation rules of the charge and current densities
are derived from some additional assumptions; typically from one of
the followings:

\begin{lyxlist}{00.00.0000}
\item [{(2a1)}] the invariance of electric charge (Jackson 1999, pp. 553--558)
\item [{(2a2)}] the current density is of form $\varrho\mathbf{u}(\mathbf{r},t)$,
where $\mathbf{u}(\mathbf{r},t)$ is a velocity field (Tolman 1949,
p. 85; M\o ller 1955, p. 140). 
\end{lyxlist}
\item [{(2b)}] The transformation of the field strengths are derived from
the transformation of $\varrho$ and $\mathbf{j}$ and from the presumption
of the covariance of the inhomogeneous Maxwell equations.
\end{lyxlist}
Unfortunately, with the only exception of (1b), none of the above
steps is completely correct. Without entering into the details, let
us mention that (2a1) and (2a2) both involve some further empirical
information about the world, which does not follow from the simple
assumption of covariance. Even in case of (1a) we must have the tacit
assumption that zero charge and current densities go to zero charge
and current densities during the transformation---otherwise the covariance
of the homogeneous Maxwell equations would not follow from the assumed
covariance of the Maxwell equations. (See points (b) and (d) in Remark~\ref{meg:kovariancia-uj})

One encounters the next major difficulty in both (1a) and (2b): neither
the homogeneous nor the inhomogeneous Maxwell equations determine
the transformation rules of the field variables uniquely; $\mathbf{E}'$
and $\mathbf{B}'$ are only determined by $\mathbf{E}$ and $\mathbf{B}$
up to an arbitrary solution of the homogeneous equations. 

Finally, let us mention a conceptual confusion that seems to be routinely
overlooked in (1c), (2a1) and (2a2). There is no such thing as a simple
relation between the scalar invariance of charge and the transformation
of charge and current densities, as is usually claimed. For example,
it is meaningless to say that 
\begin{equation}
Q=\varrho\Delta W=Q'=\varrho'\Delta W'
\end{equation}
where $\Delta W$ denotes a volume element, and
\begin{equation}
\Delta W'=\gamma\Delta W
\end{equation}
Whose charge is $Q$, which remains invariant? Whose volume is $\Delta W$
and in what sense is that volume Lorentz contracted? In another form,
in (2a2), whose velocity is $\mathbf{u}(\mathbf{r},t)$? \hfill{}
$\lrcorner$

\R\label{meg:8}In the previous remark we pointed out typical problems
in the derivations of the transformation rules\emph{ from the} \emph{covariance}
of the equations. There is however a more fundamental problem: How
do we arrive at the covariance itself? Obviously, it would be a completely
mistaken idea to regard covariance as a {}``known/verifiable property
of the equations'', because we cannot verify that the equations are
covariant against the transformations of electrodynamical quantities,
\emph{prior to} us knowing the transformations themselves against
which the equations must be covariant. Therefore, the usual claim
is that the covariance of the equations of ED against the transformations
of electrodynamical quantities---whatever these transformations are---is
\emph{implied} by the assumption that the RP holds. Strictly speaking,
as we have seen in section~\ref{sec:Mathematics-of-relativity},
this implication is not true. Covariance follows from the RP only
if $M_{\mathbf{V}}$ satisfies the additional condition \eqref{eq:RE-kov2}.
Taking into account the conceptual difficulties discussed in the next
section, it is far from obvious what $M_{\mathbf{V}}$ actually means
in ED, and whether it satisfies the required condition. 

In contrast, we have calculated the transformation rules from the
proper operational definitions of the basic electrodynamical quantities,
and have shown that the Maxwell--Lorentz equations are indeed covariant
against these transformations---\emph{independently} of the RP. In
fact, the question whether the RP holds for ED has been left open.\hfill{}
$\lrcorner$

\section{Is the RP consistent with the laws of ED in a single frame of reference?\label{sec:Is-relativity-principle}}

One might think, we simply have to verify whether the solutions of
equations (\ref{eq:MLE1})--(\ref{eq:MLE5}) satisfy condition (\ref{eq:RP-math1})
in section~\ref{sec:Mathematics-of-relativity}. Yet, this is not
possible, because we still have not determined a crucial notion in
the RP; namely, the notion of $M_{\mathbf{V}}$. This notion is however
especially problematic in case of a coupled particles + electromagnetic
field system, as the following considerations will demonstrate.

\begin{figure}
\begin{centering}
\includegraphics[width=0.6\columnwidth]{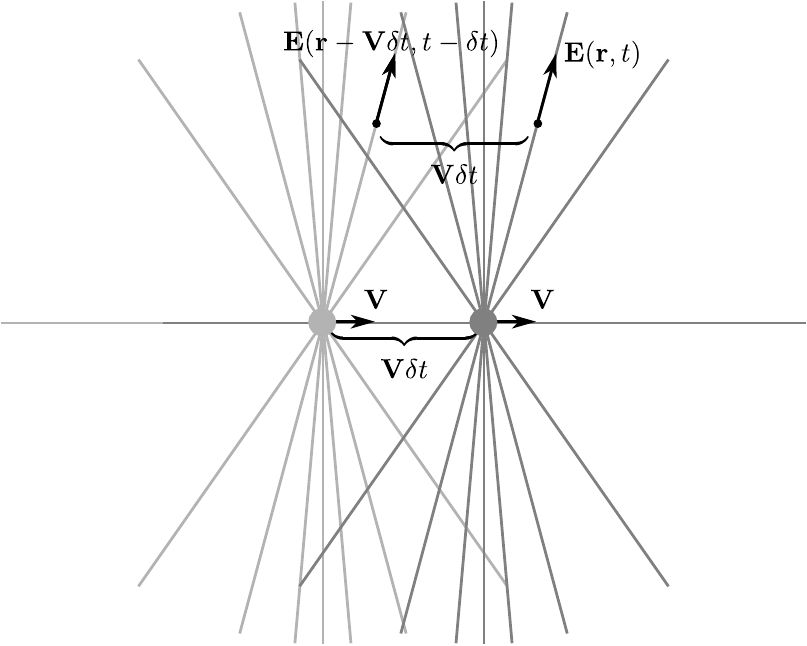}
\par\end{centering}

\caption{The stationary field of a uniformly moving point charge is in collective
motion together with the point charge \label{fig:The-stationary-field}}
\end{figure}

The reason is that a meaningful definition of $M_{\mathbf{V}}$ requires
that the solutions of equations (\ref{eq:MLE1})--(\ref{eq:MLE5})
satisfy the minimal condition (M) in Remark~\ref{meg:(M)}. Thus,
in order to make the RP applicable to ED, we need a clear answer to
the following question:
\begin{lyxlist}{00.00.0000}
\item [{(Q5)}] What meaning can be attached to the words {}``a coupled
particles + electromagnetic field system is \emph{in collective} \emph{motion}
with velocity $\mathbf{V}$'' ($\mathbf{V}=0$ included) relative
to a reference frame $K$? 
\end{lyxlist}
If this question is meaningful at all, if it is meaningful to talk
of a {}``collective motion'' of the particles + electromagnetic
field system, then it must be meaningful to talk of the instantaneous\emph{
}motion of the local parts of the system. This is no problem in case
of the particles, but we also must have a clear answer to the following
question:
\begin{lyxlist}{00.00.0000}
\item [{(Q6)}] What meaning can be attached to the words {}``the electromagnetic
field at point $\mathbf{r}$ and time $t$ \emph{is} \emph{in motion}
\emph{with some local and instantaneous velocity} $\mathbf{v}(\mathbf{r},t)$''? 
\end{lyxlist}
We can rely on what seems to be commonly accepted: According to the
\emph{application} of the RP in the derivation of electromagnetic
field of a uniformly moving point charge, the system of the moving
charged particle + its electromagnetic field \emph{qualifies} as the
system of the charged particle + its field in collective motion (Fig.~\ref{fig:The-stationary-field}).
If so, one might think, we can read off the general answer to question
(Q5): the electromagnetic field in collective motion with the point
charge of velocity $\mathbf{V}$ can be characterized by the following
condition:%
\footnote{It must be pointed out that velocity $\mathbf{V}$ conceptually differs
from the speed of light $c$. Basically, $c$ is a constant of nature
in the Maxwell--Lorentz equations, which can emerge in the solutions
of the equations; and, in some cases, it can be interpreted as the
velocity of propagation of changes in the electromagnetic field. For
example, in our case, the stationary field of a uniformly moving point
charge, in collective motion with velocity $\mathbf{V},$ can be constructed
from the superposition of retarded potentials, in which the retardation
is calculated with velocity $c$; nevertheless, the two velocities
are different concepts. To illustrate the difference, consider the
fields of a charge at rest (\ref{eq:Coulomb field}), and in motion
(\ref{eq:Coulomb-mozgo}). The speed of light $c$ plays the same
role in both cases. Both fields can be constructed from the superposition
of retarded potentials in which the retardation is calculated with
velocity $c$. Also, in both cases, a small local perturbation in
the field configuration would propagate with velocity $c$. But still,
there is a consensus to say that the system described by (\ref{eq:Coulomb field})
is at rest while the one described by (\ref{eq:Coulomb-mozgo}) is
moving with velocity $\mathbf{V}$ (together with $K'$, relative
to $K$.) A good analogy would be a Lorentz contracted moving rod:
$\mathbf{V}$ is the velocity of the rod, which differs from the speed
of sound in the rod. %
} 

\begin{eqnarray}
\mathbf{E}(\mathbf{r},t) & = & \mathbf{E}(\mathbf{r}-\mathbf{V}\delta t,t-\delta t)\label{eq:mozgomezo-0-1}\\
\mathbf{B}(\mathbf{r},t) & = & \mathbf{B}(\mathbf{r}-\mathbf{V}\delta t,t-\delta t)\label{eq:mozgomezo-0-2}
\end{eqnarray}
that is,
\begin{eqnarray}
-\partial_{t}\mathbf{E}(\mathbf{r},t) & = & \mathsf{D}\mathbf{E}(\mathbf{r},t)\mathbf{V}\label{eq:mozgomezo1}\\
-\partial_{t}\mathbf{B}(\mathbf{r},t) & = & \mathsf{D}\mathbf{B}(\mathbf{r},t)\mathbf{V}\label{eq:mozgomezo2}
\end{eqnarray}
where $\mathsf{D}\mathbf{E}(\mathbf{r},t)$ and $\mathsf{D}\mathbf{B}(\mathbf{r},t)$
denote the spatial derivative operators (Jacobians for variables $x,y$
and $z$); that is, in components:
\begin{eqnarray}
-\partial_{t}E_{x}(\mathbf{r},t) & = & V_{x}\partial_{x}E_{x}(\mathbf{r},t)+V_{y}\partial_{y}E_{x}(\mathbf{r},t)+V_{z}\partial_{z}E_{x}(\mathbf{r},t)\label{eq:elsokomponens}\\
-\partial_{t}E_{y}(\mathbf{r},t) & = & V_{x}\partial_{x}E_{y}(\mathbf{r},t)+V_{y}\partial_{y}E_{y}(\mathbf{r},t)+V_{z}\partial_{z}E_{y}(\mathbf{r},t)\label{eq:masodikkomponens}\\
 & \vdots\nonumber \\
-\partial_{t}B_{z}(\mathbf{r},t) & = & V_{x}\partial_{x}B_{z}(\mathbf{r},t)+V_{y}\partial_{y}B_{z}(\mathbf{r},t)+V_{z}\partial_{z}B_{z}(\mathbf{r},t)\label{eq:utolsokomponens}
\end{eqnarray}

Of course, if conditions (\ref{eq:mozgomezo1})--(\ref{eq:mozgomezo2})
hold for all $(\mathbf{r},t)$ then the general solution of the partial
differential equations (\ref{eq:mozgomezo1})--(\ref{eq:mozgomezo2})
has the following form:
\begin{eqnarray}
\mathbf{E}(\mathbf{r},t) & = & \mathbf{E}_{0}(\mathbf{r}-\mathbf{V}t)\label{eq:stac1}\\
\mathbf{B}(\mathbf{r},t) & = & \mathbf{B}_{0}(\mathbf{r}-\mathbf{V}t)\label{eq:stac2}
\end{eqnarray}
with some time-independent $\mathbf{E}_{0}(\mathbf{r})$ and $\mathbf{B}_{0}(\mathbf{r})$.
In other words, the field must be a stationary one, that is, a translation
of a static field with velocity $\mathbf{V}$. This is correct in
the case of a single moving point charge, provided that $\mathbf{E}_{0}(\mathbf{r})$
and $\mathbf{B}_{0}(\mathbf{r})$ are the electric and magnetic parts
of the {}``flattened'' Coulomb field \eqref{eq:Coulomb-mozgo} at
time $t_{0}$. But, (\ref{eq:stac1})--(\ref{eq:stac2}) is certainly
not the case in general; the field is not necessarily stationary. 

So, this example does not help to find a general answer to question
(Q5), but it may help to find the answer to question (Q6). For, from
(\ref{eq:mozgomezo-0-1})--(\ref{eq:mozgomezo-0-2}), it is quite
natural to say that the electromagnetic field at point $\mathbf{r}$
and time $t$ is\emph{ }moving\emph{ }with \emph{local} and \emph{instantaneous}
velocity $\mathbf{v}(\mathbf{r},t)$ if and only if
\begin{eqnarray}
\mathbf{E}(\mathbf{r},t) & = & \mathbf{E}\left(\mathbf{r}-\mathbf{v}(\mathbf{r},t)\delta t,t-\delta t\right)\label{eq:mozgomezo-0a-1}\\
\mathbf{B}(\mathbf{r},t) & = & \mathbf{B}\left(\mathbf{r}-\mathbf{v}(\mathbf{r},t)\delta t,t-\delta t\right)\label{eq:mozgomezo-0a-2}
\end{eqnarray}
are satisfied \emph{locally,} in an \emph{infinitesimally} small space
and time region at $(\mathbf{r},t)$, for infinitesimally small $\delta t$.
In other words, the equations (\ref{eq:mozgomezo1})--(\ref{eq:mozgomezo2})
must be satisfied \emph{locally} at point $(\mathbf{r},t)$ with a
local and instantaneous velocity $\mathbf{v}(\mathbf{r},t)$: 
\begin{eqnarray}
-\partial_{t}\mathbf{E}(\mathbf{r},t) & = & \mathsf{D}\mathbf{E}(\mathbf{r},t)\mathbf{v}(\mathbf{r},t)\label{eq:mozgomezo1a}\\
-\partial_{t}\mathbf{B}(\mathbf{r},t) & = & \mathsf{D}\mathbf{B}(\mathbf{r},t)\mathbf{v}(\mathbf{r},t)\label{eq:mozgomezo2a}
\end{eqnarray}

Now, if the RP, as it is believed, applies to all situations, the
concept of {}``electromagnetic field moving\emph{ }with velocity
$\mathbf{v}(\mathbf{r},t)$ at point $\mathbf{r}$ and time $t$''
must be meaningful, in other words, there must exist a local instantaneous
velocity field $\mathbf{v}(\mathbf{r},t)$ satisfying (\ref{eq:mozgomezo1a})--(\ref{eq:mozgomezo2a}),
for all possible solutions of the Maxwell--Lorentz equations. That
is, substituting an arbitrary solution of (\ref{eq:MLE1})--(\ref{eq:MLE5})
into (\ref{eq:mozgomezo1a})--(\ref{eq:mozgomezo2a}), the overdetermined
system of equations must have a solution for $\mathbf{v}(\mathbf{r},t)$. 

However, one encounters the following difficulty: 
\begin{thm}
There is a dense subset of solutions \emph{$\left(\mathbf{E}(\mathbf{r},t),\mathbf{B}(\mathbf{r},t)\right)$}
of the coupled Maxwell--Lorentz equations (\ref{eq:MLE1})--(\ref{eq:MLE5})
for which there cannot exist a local instantaneous velocity field
$\mathbf{v}(\mathbf{r},t)$ satisfying (\ref{eq:mozgomezo1a})--(\ref{eq:mozgomezo2a}).\end{thm}
\begin{proof}
The proof is almost trivial for a locus $(\mathbf{r},t)$ where there
is a charged point particle. However, in order to avoid the eventual
difficulties concerning the physical interpretation, we are providing
a proof for a point $(\mathbf{r}_{*},t_{*})$ where there is assumed
no source at all. 

Consider a solution $\left(\mathbf{r}^{b_{1}}\left(t\right),\ldots\mathbf{r}^{b_{n}}\left(t\right),\mathbf{E}(\mathbf{r},t),\mathbf{B}(\mathbf{r},t)\right)$
of the coupled Maxwell--Lorentz equations (\ref{eq:MLE1})--(\ref{eq:MLE5}),
which satisfies (\ref{eq:mozgomezo1a})--(\ref{eq:mozgomezo2a}).
At point $(\mathbf{r}_{*},t_{*})$, the following equations hold:
\begin{eqnarray}
-\partial_{t}\mathbf{E}(\mathbf{r}_{*},t_{*}) & = & \mathsf{D}\mathbf{E}(\mathbf{r}_{*},t_{*})\mathbf{v}(\mathbf{r}_{*},t_{*})\label{eq:mozgomezo1c}\\
-\partial_{t}\mathbf{B}(\mathbf{r}_{*},t_{*}) & = & \mathsf{D}\mathbf{B}(\mathbf{r}_{*},t_{*})\mathbf{v}(\mathbf{r}_{*},t_{*})\label{eq:mozgomezo2c}\\
\partial_{t}\mathbf{E}(\mathbf{r}_{*},t_{*}) & = & c^{2}\nabla\times\mathbf{B}(\mathbf{r}_{*},t_{*})\label{eq:elsomaxwell}\\
-\partial_{t}\mathbf{B}(\mathbf{r}_{*},t_{*}) & = & \nabla\times\mathbf{E}(\mathbf{r}_{*},t_{*})\label{eq:masodikmaxwell}\\
\nabla\cdot\mathbf{E}(\mathbf{r}_{*},t_{*}) & = & 0\label{eq:dive}\\
\nabla\cdot\mathbf{B}(\mathbf{r}_{*},t_{*}) & = & 0\label{eq:divb}
\end{eqnarray}
Without loss of generality we can assume---at point $\mathbf{r}_{*}$
and time $t_{*}$---that operators $\mathsf{D}\mathbf{E}(\mathbf{r}_{*},t_{*})$
and $\mathsf{D}\mathbf{B}(\mathbf{r}_{*},t_{*})$ are invertible and
$v_{z}(\mathbf{r}_{*},t_{*})\neq0$.

Now, consider a $3\times3$ matrix $J$ such that 
\begin{equation}
J=\left(\begin{array}{ccc}
\partial_{x}E_{x}(\mathbf{r}_{*},t_{*}) & J_{xy} & J_{xz}\\
\partial_{x}E_{y}(\mathbf{r}_{*},t_{*}) & \partial_{y}E_{y}(\mathbf{r}_{*},t_{*}) & \partial_{z}E_{y}(\mathbf{r}_{*},t_{*})\\
\partial_{x}E_{z}(\mathbf{r}_{*},t_{*}) & \partial_{y}E_{z}(\mathbf{r}_{*},t_{*}) & \partial_{z}E_{z}(\mathbf{r}_{*},t_{*})
\end{array}\right)\label{eq:jacobi}
\end{equation}
with

\begin{eqnarray}
J_{xy} & = & \partial_{y}E_{x}(\mathbf{r}_{*},t_{*})+\lambda\label{eq:jatek0}\\
J_{xz} & = & \partial_{z}E_{x}(\mathbf{r}_{*},t_{*})-\lambda\frac{v_{y}(\mathbf{r}_{*},t_{*})}{v_{z}(\mathbf{r}_{*},t_{*})}\label{eq:jatek}
\end{eqnarray}
by virtue of which
\begin{eqnarray}
J_{xy}v_{y}(\mathbf{r}_{*},t_{*})+J_{xz}v_{z}(\mathbf{r}_{*},t_{*}) & = & v_{y}(\mathbf{r}_{*},t_{*})\partial_{y}E_{x}(\mathbf{r}_{*},t_{*})\nonumber \\
 &  & +v_{z}(\mathbf{r}_{*},t_{*})\partial_{z}E_{x}(\mathbf{r}_{*},t_{*})
\end{eqnarray}
Therefore, $J\mathbf{v}(\mathbf{r}_{*},t_{*})=\mathsf{D}\mathbf{E}(\mathbf{r}_{*},t_{*})\mathbf{v}(\mathbf{r}_{*},t_{*})$.
There always exists a vector field $\mathbf{E}_{\lambda}^{\#}(\mathbf{r})$
such that its Jacobian matrix at point $\mathbf{r}_{*}$ is equal
to $J$. Obviously, from (\ref{eq:dive}) and (\ref{eq:jacobi}),
$\nabla\cdot\mathbf{E}_{\lambda}^{\#}(\mathbf{r}_{*})=0$. Therefore,
there exists a solution of the Maxwell--Lorentz equations, such that
the electric and magnetic fields $\mathbf{E}_{\lambda}(\mathbf{r},t)$
and $\mathbf{B}_{\lambda}(\mathbf{r},t)$ satisfy the following conditions:%
\footnote{$\mathbf{E}_{\lambda}^{\#}(\mathbf{r})$ and $\mathbf{B}_{\lambda}(\mathbf{r},t_{*})$
can be regarded as the initial configurations at time $t_{*}$; we
do not need to specify a particular choice of initial values for the
sources.%
} 
\begin{eqnarray}
\mathbf{E}_{\lambda}(\mathbf{r},t_{*}) & = & \mathbf{E}_{\lambda}^{\#}(\mathbf{r})\\
\mathbf{B}_{\lambda}(\mathbf{r},t_{*}) & = & \mathbf{B}(\mathbf{r},t_{*})
\end{eqnarray}
At $(\mathbf{r}_{*},t_{*})$, such a solution obviously satisfies
the following equations: 
\begin{eqnarray}
\partial_{t}\mathbf{E}_{\lambda}(\mathbf{r}_{*},t_{*}) & = & c^{2}\nabla\times\mathbf{B}(\mathbf{r}_{*},t_{*})\label{eq:elsomaxwell*}\\
-\partial_{t}\mathbf{B}_{\lambda}(\mathbf{r}_{*},t_{*}) & = & \nabla\times\mathbf{E}_{\lambda}^{\#}(\mathbf{r}_{*})\label{eq:masodikmaxwell*}
\end{eqnarray}
therefore 
\begin{equation}
\partial_{t}\mathbf{E}_{\lambda}(\mathbf{r}_{*},t_{*})=\partial_{t}\mathbf{E}(\mathbf{r}_{*},t_{*})\label{eq:idoderivaltakegyenlok}
\end{equation}

As a little reflection shows, if $\mathsf{D}\mathbf{E}_{\lambda}^{\#}(\mathbf{r}_{*})$,
that is $J$, happened to be not invertible, then one can choose a
\emph{smaller} $\lambda$ such that $\mathsf{D}\mathbf{E}_{\lambda}^{\#}(\mathbf{r}_{*})$
becomes invertible (due to the fact that $\mathsf{D}\mathbf{E}(\mathbf{r}_{*},t_{*})$
is invertible), and, at the same time, 
\begin{equation}
\nabla\times\mathbf{E}_{\lambda}^{\#}(\mathbf{r}_{*})\neq\nabla\times\mathbf{E}(\mathbf{r}_{*},t_{*})\label{eq:rotnem}
\end{equation}
Consequently, from \eqref{eq:idoderivaltakegyenlok} , \eqref{eq:jatek}
and \eqref{eq:mozgomezo1c} we have
\begin{equation}
-\partial_{t}\mathbf{E}_{\lambda}(\mathbf{r}_{*},t_{*})=\mathsf{D}\mathbf{E}_{\lambda}(\mathbf{r}_{*},t_{*})\mathbf{v}(\mathbf{r}_{*},t_{*})=\mathsf{D}\mathbf{E}_{\lambda}^{\#}(\mathbf{r}_{*})\mathbf{v}(\mathbf{r}_{*},t_{*})
\end{equation}
and $\mathbf{v}(\mathbf{r}_{*},t_{*})$ is uniquely determined by
this equation. On the other hand, from \eqref{eq:masodikmaxwell*}
and \eqref{eq:rotnem} we have
\begin{equation}
-\partial_{t}\mathbf{B}_{\lambda}(\mathbf{r}_{*},t_{*})\neq\mathsf{D}\mathbf{B}_{\lambda}(\mathbf{r}_{*},t_{*})\mathbf{v}(\mathbf{r}_{*},t_{*})=\mathsf{D}\mathbf{B}(\mathbf{r}_{*},t_{*})\mathbf{v}(\mathbf{r}_{*},t_{*})
\end{equation}
because $\mathsf{D}\mathbf{B}(\mathbf{r}_{*},t_{*})$ is invertible,
too. That is, for $\mathbf{E}_{\lambda}(\mathbf{r},t)$ and $\mathbf{B}_{\lambda}(\mathbf{r},t)$
there is no local and instantaneous velocity at point $\mathbf{r}_{*}$
and time $t_{*}$. 

At the same time, $\lambda$ can be arbitrary small, and 
\begin{eqnarray}
\lim_{\lambda\rightarrow0}\mathbf{E}_{\lambda}(\mathbf{r},t) & = & \mathbf{E}(\mathbf{r},t)\\
\lim_{\lambda\rightarrow0}\mathbf{B}_{\lambda}(\mathbf{r},t) & = & \mathbf{B}(\mathbf{r},t)
\end{eqnarray}
Therefore solution $\left(\mathbf{r}_{\lambda}^{b_{1}}\left(t\right),\ldots\mathbf{r}_{\lambda}^{b_{n}}\left(t\right),\mathbf{E}_{\lambda}(\mathbf{r},t),\mathbf{B}_{\lambda}(\mathbf{r},t)\right)$
can fall into an arbitrary small neighborhood of $\left(\mathbf{r}^{b_{1}}\left(t\right),\ldots\mathbf{r}^{b_{n}}\left(t\right),\mathbf{E}(\mathbf{r},t),\mathbf{B}(\mathbf{r},t)\right)$.
\end{proof}
Thus, the meaning of the concept of {}``electromagnetic field moving\emph{
}with velocity $\mathbf{v}(\mathbf{r},t)$ at point $\mathbf{r}$
and time $t$'', that we obtained by generalizing the example of
the stationary field of a uniformly moving charge, is untenable. Perhaps
there is no other available rational meaning of this concept. In any
event, lacking a better suggestion, we must conclude that the question
whether the relativity principle generally holds in classical electrodynamics
remains not only unanswered, but even ununderstood.

\section*{Acknowledgment}

The research was partly supported by the OTKA Foundation, No.~K 68043.

\section*{References}

~
\begin{lyxlist}{00.00.0000}
\item [{Bell,~J.S.~(1987):}] How to teach special relativity, in \emph{Speakable
and unspeakable in quantum mechanics}. Cambridge, Cambridge University
Press.
\item [{Einstein,~A~(1905):}] \foreignlanguage{ngerman}{Zur Elektrodynamik
bewegter Körper, \emph{Annalen der Physik}} \textbf{17}, 891. (On
the Electrodynamics of Moving Bodies, in H. A. Lorentz et al.,\emph{
The principle of relativity: a collection of original memoirs on the
special and general theory of relativity. }London, Methuen and Company
1923)
\item [{Frisch,~M.~(2005):}] \emph{Inconsistency, Asymmetry, and Non-Locality},
Oxford, Oxford University Press. 
\item [{Georgiou,~A.~(1969):}] Special relativity and thermodynamics,
\emph{Proc. Comb. Phil. Soc.} \textbf{66}, 423. 
\item [{Jackson,~J.D.~(1999):}] \emph{Classical Electrodynamics (Third
edition).} Hoboken (NJ), John Wiley \& Sons.
\item [{Jammer,~M.~(2000):}] \emph{Concepts of Mass in Contemporary Physics
and Philosophy. }Princeton, Princeton University Press.
\item [{M\o ller~C.~(1955):}] \emph{The Theory of Relativity.} Oxford,
Clarendon Press. 
\item [{Muller,~F.~(2007):}] Inconsistency in Classical Electrodynamics?,
\emph{Philosophy of Science} \textbf{74}, pp. 253-277. 
\item [{Jánossy,~L.~(1971):}] \emph{Theory of Relativity Based On Physical
Reality}. Budapest, \inputencoding{latin2}\foreignlanguage{magyar}{Akadémiai
Kiadó}\inputencoding{latin9}.
\item [{Quine,~W.V.O.~(1951):}] Two Dogmas of Empiricism, \emph{The Philosophical
Review} \textbf{60}, pp. 20--43.
\item [{Reichenbach,~H.~(1965):}] \emph{The Theory of Relativity and
A Priori Knowledge. }Berkeley and Los Angeles, University of California
Press.
\item [{Rohrlich,~F.~(2007):}] \emph{Classical Charged Particles. }Singapore,
World Scientific\emph{.}
\item [{Szabó,~L.E.~(2004):}] On the meaning of Lorentz covariance, \emph{Foundations
of Physics Letters} \textbf{17}, pp. 479--496.
\item [{Szabó,~L.E.~(2009):}] Empirical foundation of space and time,\emph{
}in M. Suárez et al. (eds.), \emph{EPSA Philosophical Issues in the
Sciences: a Launch of the European Philosophy of Science Association}.
Berlin, Springer.
\item [{Tolman,~R.C.~(1949):}] \emph{Relativity, Thermodynamics and Cosmology.}
Oxford, Clarendon Press.\end{lyxlist}

\end{document}